\documentclass[preprint2]{aastex}

\usepackage{graphicx}
\usepackage{epsfig}
\usepackage{lscape}
\usepackage{graphicx}
\usepackage{amsmath}%
\usepackage[english]{babel}
\usepackage[titletoc,title]{appendix}
\usepackage{txfonts}
\usepackage{times}
\usepackage{natbib}
\begin{document}
\title{Constraints on Dynamical Dark Energy Models from the Abundance of Massive Galaxies at High Redshifts} 
\author{N. Menci$^1$,  A. Grazian$^2$, M. Castellano$^1$, P. Santini$^1$, E. Giallongo$^1$, A. Lamastra$^1$, F. Fortuni$^1$, A. Fontana$^1$, E. Merlin$^1$, T. Wang$^{3}$, D. Elbaz$^{4}$,  N.G. Sanchez$^5$.
}
\affil{$^1$INAF - Osservatorio Astronomico di Roma, via Frascati 33, I-00078 Monteporzio, Italy}
\affil{$^2$INAF--Osservatorio Astronomico di Padova, Vicolo dell'Osservatorio 5, I-35122, Padova, Italy}
\affil{$^3$National Astronomical Observatory of Japan, Mitaka, Tokyo, Japan}
\affil{$^4$AIM, CEA, CNRS, Université Paris-Saclay, Université Paris Diderot, Sorbonne Paris Cité, Gif-sur-Yvette, France}
\affil{$^5$LERMA, CNRS UMR 8112, 61, Observatoire de Paris PSL, Sorbonne Universit\'es, UPMC Univ.\ Paris 6, 61 Avenue de l'Observatoire, F-75014 Paris, France}

\begin{abstract} 
We compare the maximal abundance of massive systems predicted in different dynamical dark energy (DDE) models at high redshifts $z \approx 4-7$ with the measured abundance of the most massive galaxies observed to be already in place at such redshifts. The aim is to derive constraints for the evolution of the dark energy equation of state parameter $w$ which are complementary to existing probes.  We adopt the standard parametrization for the DDE evolution in terms of the local value $w_0$ and of the look-back time derivative $w_a$ of the equation of state. We derive constraints on combinations of $(w_0,  w_a)$ in the different DDE models by using three different, independent probes: (i) the observed stellar mass function of massive objects at $z \geq 6$ derived from the CANDELS survey; (ii) the estimated volume density of massive halos derived from the observation of massive, star-forming galaxies detected in the submillimeter range at $z \approx 4$; (iii) The rareness of he most massive system (estimated gas mass exceeding $3\cdot10^{11} M_{\odot}$) observed to be in place at $z \approx 7$, a far-infrared-luminous object recently detected in the South Pole Telescope (SPT) survey. Finally, we show that the combination of our results from the three above probes {\it excludes a sizable fraction} of the DDE parameter space $w_a \gtrsim -3/4 - (w_0 + 3/2)$ presently allowed (or even favored) by existing probes.
\end{abstract}

\section{Introduction} 
The current theory of structure formation envisages all cosmic structures to form from the collapse and the growth of initially tiny density perturbations of  dark matter (DM) density field in a Universe characterized by an accelerated expansion. Such an acceleration indicates that the dominant component (with density parameter $\Omega_{\Lambda}\approx 0.7$) of the cosmic fluid must be constituted by some form of dark energy (DE), with equation-of-state parameter $w\equiv p/\rho\leq -1/3$. Although the nature of such a component  remains unknown, the simplest model assumes DE to be connected with the vacuum energy, the so-called cosmological constant, with equation-of-state parameter $w=-1$. When coupled with the assumption that DM is constituted by non-relativistic particles at decoupling, such a scenario  leads to the $\Lambda$CDM standard cosmological model. 
 
While measurements of the Cosmic Microwave Background (CMB) have provided a first, strong confirmation of such a scenario, 
 tensions are recently emerging (mostly between measurements related to the early and late universe): these include the inadequacy of the model in providing a perfect fit
to the Planck CMB temperature and polarization angular spectra (see, e.g., Addison et al. 2016), the  discrepancy between the combined values of 
the power spectrum normalization $\sigma_8$ and matter density parameter $\Omega_M$ 
 derived by Planck with respect to those derived from cosmic shear surveys such as CFHTLenS (Heymans et al. 2012) and KiDS-450 (Hildebrandt et al. 2017),  and - most of all - the tension (at more than 3-$\sigma$ confidence level) in the Hubble constant $H_0$ between the values derived from Planck ($h\approx 0.67$, in units km\,s$^{-1}$\,Mpc$^{-1}$) and those obtained  from local luminosity distance measurements ($h\approx 0.74$, see Riess 2019 and references therein). 
 
 Such tensions have stimulated an extended effort toward the investigation of more complex cosmological models.   
 One of the simplest physical alternatives is constituted by a DE with time-dependent equation of state (dynamical dark energy, DDE). In fact,  this kind of scenario constitutes a possible solution to the above mentioned tension between the values of the Hubble constant derived from local indicators and from the CMB (Di Valentino, Melchiorri, Linder, Silk 2017; Pan, Yang, Di Valentino et al. 2019).  In addition, a constant equation of state is not expected in physically motivated scenarios in which 
 DE  originates from a scalar ''quintessence'' field $\phi$ (Peebles, Ratra 1988; Caldwell, Dave, Steinhardt 1998; Sahni, Wang, 2000; Copeland, Sahni,  Tsujikawa,  2006; Frieman, Turner, Huterer 2008 ) evolving in a potential $V(\phi)$. In fact, its pressure $p_{\phi}=\dot \phi-V(\phi)$ and energy density $\rho_{\phi}=\dot \phi+V(\phi)$ lead to a DE with a time-evolving equation of state  parameter $w\equiv p_{\phi}/\rho_{\phi}$.
 Parametrizing the evolution of $w$ with the expansion factor $a$ as  $w(a)=w_0+w_a(1-a)$ (Chevallier and Polarski 2001, Linder 2003) the dynamics of such models can be related to different combinations ($w_0$, $w_a$) (see Caldwell and Linder 2005; Barger Guarnaccia, Marfatia 2006; Linder 2006). E.g., "thawing" models (Scherrer 2008; Chiba 2009; Gupta, Rangarajan, Sen 2015) are characterized by $w$ growing with time starting from $w=-1$ in the early universe, while in "freezing" models (Chiba 2006; Scherrer 2006; Sahlen, Liddle, Parkinson 2007) $w$ decreases with $a$ approaching a cosmological constant value $w=-1$.  
 
Present observational constraints on the DDE parameter space $w_0-w_a$ (see, e.g., Zhai et al. 2017) provide contrasting results. On the one hand, results coming  from the CMB power spectrum (Ade et al. 2016) and weak lensing tomography (Massey et al. 2007,  Amara and Refregier 2007; see Refregier 2003 for a review) - although leaving unconstrained a relatively large volume of the DDE parameter space  $w_0-w_a$ - disfavor combinations with $w_0\lesssim -1.5$ and positive values of $w_a\gtrsim 0.7$ (especially when combined with constraints from baryonic acoustic oscillations and from 
type-Ia Supernovae, see Ade et al. 2016; Scolnic 2018; di Valentino et al. 2017 and references therein), as well as most of the combinations $w_0\gtrsim -1$ and $w_a\gtrsim 0.5$.   On the other hand, the recent determination of the Hubble diagram of quasars 
 in the range $0.5\leq z\leq 5.5$  (Risaliti and Lusso 2019) favors large values $w_a\gtrsim 0$ with negative values of $w_0<-1$, with a deviation from the $\Lambda$CDM model emerging with a statistical significance of $4\sigma$. 
This measurement is based on quasar distances estimated from the ratio between their X-ray and ultraviolet emission. Although  
 it strongly relies on the assumed invariance with redshift of the X-ray-to-ultraviolet ratio, the excellent agreement with the Hubble diagram derived from type-Ia Supernovae in the overlapping range $0.5\leq z\leq 1.5$ and the evidence of non-evolving UV and X-ray spectral properties strongly argue for the reliability of such results.  

In such a context, a further, independent  probe for the nature and evolution of DE is constituted by the evolution of the galaxy population over cosmic time. In fact, the inverse dependence of the amplitude of initial density perturbation on the mass scale (measured from fluctuations of the CMB, see, e.g., Tegmark and Zaldarriaga 2002, 2009; Aghanim et al. 2019) 
implies that the formation of galactic DM halos proceeds bottom-up. Although the physics of baryons assembling into the DM halos constitutes a complex issue, a solid consequence of the above scenario is that - in  any specific adopted  cosmological model - large-mass DM haloes must become progressively rarer with increasing redshift. Thus, viable cosmological models must allow for an evolution of the 
initial density perturbations fast enough to match the abundance of massive galaxies  observed to be in place early on in history of the Universe. 
Indeed, several observations concerning massive galaxies at high redshifts are already challenging the canonical 
  $\Lambda$CDM cosmological model. E.g., several authors (see, e.g., Hildebrandt et al. 2009; Lee et al. 2012; Caputi et al. 2015; Finkelstein et al. 2015; Merlin et al. 2019; see also Wang et al. 2019) have enlightened the tension between the expected evolution of the DM halo mass function and the observed galaxy luminosity and mass functions at $z\gtrsim 4$. While the present understanding of the baryonic processes leading to gas condensation and to star formation in DM halos struggles in describing the  rapid evolution of the star formation needed to match the observed abundance of massive galaxies (see Steinhardt et al. 2016), an enhanced   efficiency in converting baryons into stars at high redshifts could still allow for consistency between the $\Lambda$CDM predictions and the     observed number density of luminous, massive galaxies  (Behroozi et al. 2013; Behroozi and Silk 2015; Finkelstein et al. 2015; Sun and Furlanetto 2016; Moster et al. 2018, Behroozi and Silk 2018). 

The above degeneracy between baryonic effects and cosmology in determining the expected abundance of luminous  galaxies can be bypassed by noticing that the ratio of galaxy baryonic components (stellar mass or gas mass) to DM halo mass has an absolute maximum at the cosmic baryon fraction  $f_b\approx 0.16$  (Aghanim et al., Planck Collaboration,  2018). In fact, the observed abundance of galaxies with large mass in the baryonic component $M_b$  places a lower limit on the abundance of DM  haloes with masses $M\geq M_b/f_b\approx 6.3\,M_b$. Such a constrain can be used to rule out cosmological models which do not allow for a sufficiently rapid growth of galactic DM halos. In fact, an observed abundance $\phi_{obs}(M_b,z)$ would rule out any cosmological models  predicting a number density of DM haloes  $\phi(M\geq M_b/f_b,z)\leq \phi_{obs}(M_b,z)$ independently of the details of the complex baryon physics involved in the galaxy formation process. 

In this paper we apply such a probe to cosmological models based on dynamical dark energy (DDE). We compare the maximal abundance of massive galaxies predicted in different DDE models at high redshifts with the measured  abundance of the most massive systems observed to be already in place at the same redshifts. 

The plan of the paper is as follows: \\
$\bullet$ in Sect. 2 we present the method adopted to derive the halo mass function in different DDE models, 
and how we compute the basic quantities that we will compare with observational data. \\
$\bullet$ Such a comparison is performed in Sect. 3 for three different  observations concerning the abundance of massive objects at high redshift: we first compare the observed stellar mass function of massive objects at $z\geq 6$ derived from the CANDELS survey with the halo mass function predicted in different DDE models (Sect. 3.1), 
 deriving  exclusion plots in the DDE parameter space $w_0-w_a$. In Sect. 3.2 we perform a comparison with the estimated volume density of massive halos derived from the observation of massive, star-forming galaxies detected in the submillimeter at $z\gtrsim 4$ (Wang et al. 2019), which are expected to reside in the most massive DM haloes at their redshift. Finally, in sect. 3.3 we consider the most massive object (estimated gas mass exceeding  $3\cdot 10^{11}\,M_{\odot}$) in place  at $z\approx 7$ recently detected in the SPT survey, and we derive (adopting the most conservative assumptions for the baryon-to-DM ratio) the probability for such a massive object to be present in the area covered by SPT for different DDE models for different combinations ($w_0, w_a)$. The constraints derived from the combination of the above observations are shown in Sect. 3.4. \\
$\bullet$ The final Sect. 4 is devoted to discussion and conclusions.
 
\section{Method} 
To compute the expected abundance of DM haloes in different DDE we adopt the canonical Press and Schechter approach, which relates the number of halos 
of mass $M$ at redshift $z$ to the overdense regions of the linear density field with density contrast $\delta$ over a background average density 
$\overline{\rho}$.
Successful models predict the halo mass function $\phi$ (i.e., the number of virialized DM haloes with mass in the range $M-M+dM$ per unit volume) to take the form
\begin{equation}
\phi(M)={dN\over dM}= {\overline{\rho}\over M^{2}}\,{d ln\, \nu\over d\,\,ln M}\,f(\nu)
\end{equation}
Here $\nu=\delta_c/\sigma(M,z)$ where $\delta_c$ corresponds  to the critical linear overdensity (equal to 1.69 in the $\Lambda$CDM scenario) and 
$\sigma(M ,z)$ is the variance of the linear density field smoothed on the scale 
$R=[3M/4\pi\overline{\rho}]^{1/3}$, and evolving with time according to the linear growth factor $D(z)$ of density perturbations.  
The function $f(\nu)$ is universal to the changes in redshift and cosmology. 
In the original Press-Schechter theory (Press and Schechter 1974) and in excursion set theory  (Bond et al. 1991) the function
$f$ takes the form 
$f_{PS}=(\sqrt{2/ \pi})^{1/2} {\nu}\,exp(-\nu^{2}/2)$ appropriate for spherical collapse. More recent approaches (Sheth and Tormen 1999, Jenkins et al. 2001; Warren et al. 2006; Tinker et al. 2008) provided more accurate forms that have been extensively tested against N-body simulations. 
Here we adopt the form given by Sheth and Tormen (1999): 
\begin{equation}
f(\nu)=2\,A\,\Bigg({1 \over \nu^{'2q}}+1\Bigg)\,  {\nu'^{2}\over 2\,\pi}     e^{-\nu'^{2}/2}. 
\end{equation}
with $\nu'=\sqrt{a}\nu$, $a=0.71$, $q=0.3$. The normalization factor (ensuring that the integral of $f(\nu)$ gives unity) is $A=0.32$. 
Corresponding theoretical advances (see, e.g., Sheth et al. 2001, Maggiore and Riotto 2010, 
Corasaniti and Achitouv, 2011a; Achitouv and Corasaniti 2012) 
 have enlightened the physical meaning of the coefficients $a$ and $q$  in the framework of the excursion set theory (the normalization $A$ is not an independent parameter since its value is obtained requiring that integral of $f(\nu)$ gives unity).  
 This enables to map the computation of the mass function in terms of  a first-passage process of a random walk across a barrier. 
 While a constant barrier corresponds to spherical collapse, 
 the above authors showed that  a drifting, diffusive barrier as a function of $\sigma$ accounts not only for the non-spherical form of the collapse, but also for more complex aspects of the underlying dynamics. The drift and diffusion coefficient of the barrier determine 
 the coefficients $q$ and $a$ in eq. 2, which are then related to the collapse process of halos. 

The universality of the above mass function and of the coefficients $a$, and $q$ for different cosmologies has been studied by various authors, starting from Sheth et al.  (2001). These authors have tested  the above expression against N-body simulations for a variety of CDM cosmologies including the cases of a critical Universe ($\Omega_M=1$, $\Omega_{\Lambda}=0$, $h=0.5$), of an open Universe ( $\Omega_M=0,3$, $\Omega_{\Lambda}=0$, $h=0.7$), and the $\Lambda$CDM case ($\Omega_M=0.3$, $\Omega_{\Lambda}=0, 7$, $h=0.7$). More recently, Despali et al. (2016) have tested the above  mass function against the SBARBINE set of  N-body simulations for a variety of combinations of $\Omega_M$ (ranging from 0.2 to 0.4)  and $\Omega_{\Lambda}$ (ranging from 0.6 to 0.8), finding an excellent agreement with simulations when the same definition of halos (based on the virial   value) is adopted, for the same, fixed set of parameters. 
These authors concluded that - with the proper definition of halo - the Sheth \& Tormen mass function is universal as a function of redshift and cosmology.  In all explored cosmologies the 
  deviations from the mass function that we adopt are shown to be within percent value. For non-standard cosmologies, 
  Achitouv, Wagner, Weller, Rasera (2014) explored a Ratra-Peebles quintessence model of Dark Energy in addition to the standard $\Lambda$CDM. They found that the  
  abundance of dark matter halos described by a drifting diffusive barrier matches the 
  results of N-body simulations to within 5\% for all explored cosmologies. Most important, the parameters defining the diffusion of the barrier (determining the coefficients $q$ and $a$) change by less than 5 \% when passing from $\Lambda$CDM to the quintessence cosmology for the large  masses $M\geq 10^{11}\,M_{\odot}$ relevant to this paper. 
  Motivated by the above results, in the rest of the paper we retain the same form 
  of the mass function for all DDE cosmologies. We have verified that allowing for a   5\% uncertainty in the 
  mass function in eq. (2) does not change our results appreciably.

As for the threshold $\delta_c$, we notice that in principle this 
depends weakly on cosmology. Here we shall adopt the conservative value $\delta_c=1.65$ for all DDE models. This constitutes a lower bound for the possible values taken in different DDE cosmologies (Mainini, Maccio, Bonometto, Klypin 2003; Pace, Waizmann, Bartelmann 2010) thus maximizing the predicted abundance of massive DM halos. 

The mass function $dN/dM$ (eq. 1) allows us to compute the 
expected number of galaxies in a given region of mass and redshift over a given fraction of the sky, $f_{sky}$ as 
\begin{equation}
N=f_{sky}\,\int\,dz\,  {dV\over dz}\,\int dM \, {dN\over dM} 
\end{equation}
where the $z$ and $M$ integrals are over the region of the $(M, z)$ plane being considered. 

The above expressions depends on the assumed power spectrum of perturbations (determining the dependence of $\sigma$ on the mass $M$) and on cosmology, which affects the volume element $dV/dz$ and the growth factor of perturbations $D(z)$. 

For the first we adopt the CDM form (Bardeen et al. 1986), which has long been known to provide an excellent match to a wide set of observational data (see Tegmark and Zaldarriaga 2002;  Tegmark and Zaldarriaga 2009;  Hlozek et al. 2012). 
In this paper we adopt a top-hat filter in real space to relate the variance $\sigma(M)$ to the power spectrum $P(k)$. 
As explained in detail in existing papers (see, e.g., Benson et al. 2012, Schneider 2013) for CDM power spectra, which are essentially pseudo power-laws with a slowly varying slope, the shape of the
 $\sigma(M)$ varies little with the choice of filter function. In any case, 
 even in cosmologies with a small scale cut-off (such as those of Warm Dark Matter models), the variance $\sigma(M)$ depends on the filter choice only at small masses $M\lesssim 10^{9}\,M_{\odot}$, while in this paper we focus large masses $M\geq 10^{10}\,M_{\odot}$ for which the variance is in practice independent of the filter choice (see Benson et al. 2013). 

The dependence on cosmology (and in particular on the DE equation of state) constitutes our main focus here. In the present paper, we follow the approach adopted in Lamastra et al. (2012), to which we refer for further details. Here we summarize the key points. 
 
We assume a spatially flat, homogeneous and isotropic universe filled by non-relativistic matter plus a dark energy component.
For our analysis, we  use the Chevallier-Polarski-Linder parametrization (Chevallier and Polarski 2001, Linder 2003) to describe the evolution in terms of the scale factor $a$ (normalized to unity at the present cosmic time): 
\begin{equation}\label{w}
w(a)=w_0+w_a(1-a)=w_0+w_a\,{z\over 1+z}
\end{equation}
where the parameter $w_0$ represents the value of $w$ at the present epoch, while $w_a$ corresponds to its look-back time variation $w_a= - \,dw/da$. In the above parametrization, the standard $\Lambda$CDM cosmology corresponds to $w_0=-1$ and $w_a=0$. 
Using this parametrization the cosmic expansion described by $H=\dot a/a$ given by
\begin{equation}\label{}
E(z)\equiv H/H_{0}=[\Omega_{M}\,a^{-3}+\Omega_{\Lambda}a^{-3(1+w_0+w_a)}e^{3w_a(a-1)} ]^{1/2}.
\end{equation}
The above equation also yields the line-of-sight comoving distance corresponding to a distant object at redshift $z$ in any DE model:
\begin{equation}\label{htz}
\chi(z)={c\over H_0}\,\int_{0}^{z} \frac{dz'}{E(z')}.
\end{equation}
This enters the expression for the luminosity and angular distances, and for the volume element $dV/dz$ (see, e.g., Weinberg 1972). 
In the following we will indicate as $V_{w_0,w_a}$ the cosmic volume computed for DDE cosmologies, while $V_{\Lambda}$ is the 
 same quantity computed in the case $w_0=-1$ and $w_a=0$ (cosmological constant). 

As for the growth factor, its expression in  the $\Lambda$CDM case is given by Carroll, Press, Turner (1992) in the form :
\begin{equation}
\delta(a)=\frac{5 \Omega_M}{2a}\frac{da}{d\tau}\int_0^a \left(\frac{da^{'}}{d\tau}\right)^{-3}da^{'}
\end{equation}
where $\tau=H_0\,t$. For the DDE models we use the parametrization to the solution given in Linder (2005):
\begin{equation}\label{d_par}
\frac{\delta(a)}{a}=exp\left(\int_{0}^{a}[\Omega (a)^{\gamma}-1]dlna\right) 
\end{equation}
where  $\Omega (a)=\Omega_M a^{-3}/(H(a)/H_{0})^2$, and $\gamma$  is the growth index, given by the fitting formula  (Linder 2005):
\begin{eqnarray} \label{gamma}
\gamma=0.55+0.05(1+w(z=1)) & w(z=1)\geq -1   \nonumber\ \\ 
\gamma=0.55+0.02(1+w(z=1)) & w(z=1)< -1  \,.
\end{eqnarray}
This parametrization reproduces the behaviour of the growth factor to within 0.1\%-0.5\% accuracy for a wide variety of dark energy cosmologies (Linder 2005, Linder and Cahn 2007)  and allows for a rapid scanning of the parameter space of DDE models. 
We  normalize the amplitude of perturbations $\delta$ 
in terms of $\sigma_8$, the the local ($z=0$) variance of the density field smoothed over regions of 8 $h^{-1}$ Mpc. 
  Present cosmological constraints based on Planck data, baryonic acoustic oscillations and type-Ia supernovae yield  $\sigma_8=0.8$   for the $\Lambda$CDM cosmology (Ade et al. 2016); such a value vary by $\approx 2$ \% when different combinations  ($w_0, w_a$)  are assumed  (Ade et al. 2016; Di Valentino 2017; Mehrabi et al. 2018).
 
Both the distance relation (eq. 6) and the growth factors (eq. 8) of DDE models deviate mildly from the cosmological constant case when the equation of state is negatively evolving  with redshift $w_{a}$ $<$0, while models with  $w_{a}$ $>$0 yield growth factors and cosmic times lower than those predicted in the  $\Lambda$CDM case. This is because in the $w_{a}$ $>$0 models  the influence of DE at early times is strong even at high redshift (see eq. (5)), yielding shorter ages and implying a delay in the growth of DM perturbations compared to the $\Lambda$CDM case (see fig. 1 in Lamastra et al. 2012). 

The impact of the above effects on the predicted abundance of halos (eq. 1)  is  illustrated in fig. 1. We show the
 DM mass function derived from eq. (1) in  selected DDE cases with four different values of $w_a$ (0 , 0.5,  0.9,  1.1) and fixed $w_0=-1$. For a prompt comparison, we have also shown the mass distribution of galaxies  (with the $2-\sigma$ errorbar) corresponding to the stellar mass function by Grazian et al. (2015), assuming that all baryons are converted into stars, i.e.,  $M_*/M=f_b$, to convert the observed stellar mass $M_*$ to the halo DM mass $M$ (see Sect. 3.1 for a detailed comparison).

 While the  mass function of observed galaxies is expected to be lower than the halo mass function due to the 
 inefficient conversion of baryons into observable stellar mass, luminous galaxies cannot outnumber their host DM halos. Thus, the DM mass function in the figure should be considered as {\it upper limits} for the mass distribution of observed galaxies.

\vspace{-0.4cm}
\begin{center}
\scalebox{0.38}[0.38]{\rotatebox{0}{\includegraphics{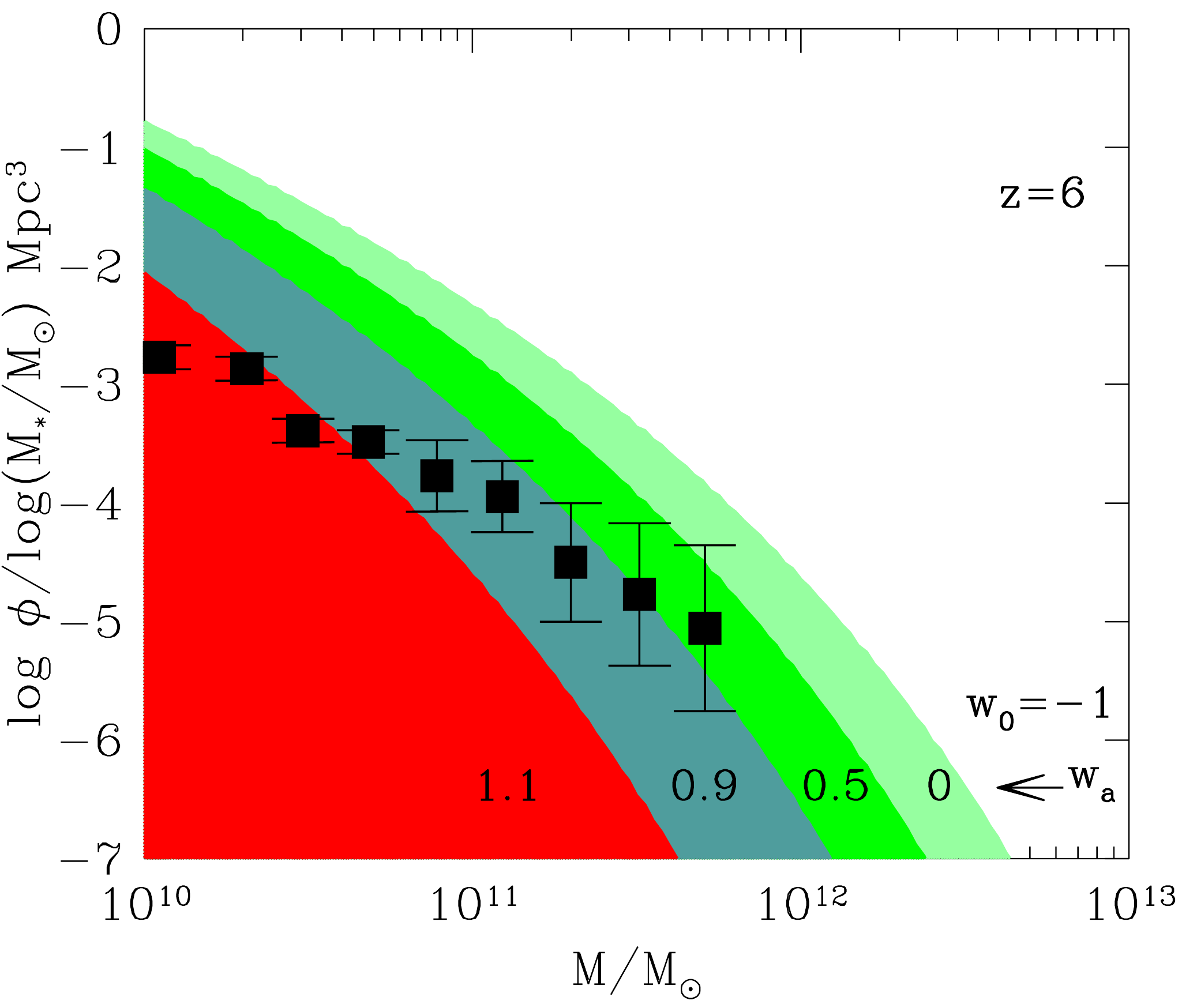}}}
\end{center}
\vspace{-0.3cm }
{\footnotesize Fig. 1. The halo mass function (solid lines) at $z=6$ predicted by different DDE models with $w_0=-1$ and $w_a=0$ , 0.5,  0.9,  1.1 (from right to left, light green, green, dark green, and red colors, respectively). The black dots correspond to stellar mass function measured by Grazian et al. (2015) at $z=6$; for the sake of simplicity, in this plot a conversion factor $M_*/M=f_b$ has been assumed to assign a DM mass to the stellar mass of the measured data point.  We have shaded the region below each curve to stress that the DM halo mass function constitutes an upper limit for the stellar mass function. In any assumed DDE scenario,  the stellar mass function must entirely lay within the 
corresponding shaded region.}
\vspace{0.1cm }

The rapid, exponential decline of the  DM mass function at large masses (eqs. 1 and 2) results into a large sensitivity of the predicted abundance of massive DM halos on the growth factor $D(z)$ corresponding to the different considered DDE models. As a result, in DDE models with $w_0=-1$ and $w_a\geq 1.1$ the maximal abundance of massive DM halos is too low (deviation larger than $2-\sigma$) to account for the observed number density of massive galaxies. 
 
The example above shows that selecting  extremely massive objects at high redshift is essential to provide constraints on  DDE models. 
Thus, in the following sections, we will compare DDE predictions with different observations concerning the most massive objects already in place at high redshifts. 

\section{Results} 

Here we compare the abundance of DM halos predicted by DDE models with different observations.  When computing the mass function (eq. 1), we assume a matter density parameter $\Omega_M=0.31$, a baryon density parameter $\Omega_b=0.045$, corresponding to the values that provide the best fit to CMB data when $w_0$ and $w_a$ are allowed to vary (see Di Valentino et al. 2017); similar results are obtained if we convolve our predictions with a Gaussian uncertainty distribution centered on the above values and with variance $\sigma_{\Omega_M}=0.02$ and $\sigma_{\Omega_b}=0.005$, respectively. 
For the Hubble constant we take the value $h=0.7$ although the best fit values to CMB (in combination with  other probes)  vary in the range $0.67\leq h\leq 0.74$ when $w_0$ and $w_a$ are allowed to vary (Di Valentino 2017). 
In fact,  the final constraints we obtain 
on the $w_0-w_a$ plane are weakly dependent on $H_0$ for the
massive systems redshift range considered   $z \approx 4-7$:
for any given $w_0$, varying $H_0$ in the above range changes our constraints on 
$w_a$ by less than 2\%. 
 
\subsection{The Stellar Mass function at $z=6$ from CANDELS}

We first compare with the observed stellar mass distribution of massive, distant galaxies. Since stellar mass is a time-integrated quantity, it is less sensitive to the details of the star formation history and can be more easily related to the DM mass of the host halo. However,  an extended wavelength coverage is essential for estimating stellar masses from SED (spectral energy distribution) fitting, while measuring the abundance of massive, rare galaxies requires a 
combination of survey volume and depth. The CANDELS project (Koekemoer et al. 2011; Grogin et al. 2011)
takes advantage of the optical/near-infrared/mid-infrared imaging provided by Hubble Space Telescope (HST), Spitzer, and the Very Large Telescope,  and provides an ideal data set to base on for such a measurement. Here we use the high redshift ($z=5.5-6.5$) mass function derived by Grazian et al. (2015),  who used a spectral-fitting technique to derive stellar masses for a galaxy sample with high-quality photometric redshifts based on the CANDELS-UDS, GOODS-South, and HUDF fields. 
The high redshifts we are considering ensure that at the largest  masses $M_*\approx10^{11}\,M_{\odot}$ probed by observations, the mass functions predicted by the different  DDE models are in the full exponential regime, and are steep enough to make the comparison with the observed number density discriminant for the different DDE models (see Sect. 2 and fig. 1) with a high ($2-\sigma$) confidence level, as we show below. 
 To take into proper account the uncertainties related to the stellar mass measurements (in turn depending on age, dust extinction, metallicity, star formation history), and to  photometric redshifts,  star formation histories,  cosmic variance and Poissonian statistical fluctuations, we have run a Monte Carlo simulation, specific to the adopted data set. This allows us to derive, for any chosen stellar mass bin, the whole probability distribution functions $p(\phi_{obs})$ (PDF hereafter) of measuring an abundance $\phi_{obs}$.

We associate the stellar mass $M_*$ to the host halo DM mass $M$ 
using the relation $M_*=F\,f_b\,M$, where $F$ describes the efficiency of baryon conversion into stars. While $F=1$ corresponds to 
the complete conversion, the different processes (gradual gas cooling gas ejection, stellar feedback) taking place in galaxies 
limit $F$ to lower values. In fact, the standard conversion for $\Lambda$CDM derived from abundance matching techniques (see, e.g., Behroozi \& Silk 2018) yields values $F\approx 0.25$. Such a value cannot be safely considered as a baseline for generic DDE models, since the stellar masses derived from abundance matching assume a $\Lambda$CDM halo mass function. However, we can study the effectiveness of baryon conversion into stars using hydrodynamical N-Body simulations, since the physics of such a conversion is expected to weakly depend on the background cosmology. To this aim we analyzed the public release of three simulations: 
the  Illustris simulations (Springel, 2010; Genel et al., 2014; Vogelsberger et al., 2014a,b), its updated version (IllustrisTNG, Weinberger et al. 2017;  Pillepich et al. 2018), and the EAGLE simulation (Schaye 2015). For  Illustris TNG  we considered the highest resolution version of the largest and medium volume realizations (TNG300 and TNG100). We computed the conversion efficiency  $F=(M_*/M)/f_b$ from the ratio between the DM mass $M$ of each sub-halo in the simulations and the stellar content $M_*$ associated to the considered sub-halo,  
finding that $F=0.5$ constitutes an effective, conservative upper limit for such a quantity, since (in all simulations) no massive  ($M_*\geq 3\cdot 10^{10}\,M_{\odot}$)  galaxies have been found with $F\geq 0.5$ at $z =5.5-6.5$.

Then, we consider a  grid of DDE models characterized by different combinations $(w_0, w_a)$. For each combination $(w_0, w_a)$ we first correct  the observed abundances $\phi_{obs}$ with the volume factor $f_{Vol}=V_{\Lambda}/V_{w_0,w_a}$ (computed in the redshift range $z=5.5-6.5$) to account for the fact that the mass function given in Grazian et al. (2015) have been derived  assuming a $\Lambda$CDM cosmology. Analogously, we must take into account  that the stellar masses measured by Grazian et al. (2015) have been inferred from luminosities assuming a $\Lambda$CDM cosmology to convert  observed fluxes into luminosities.  Thus, for each combination  $(w_0, w_a)$ we must correct the masses $M_*$ measured by Grazian et al. (2015) 
 by a factor $f_{lum}=D^{2}_{L,w_0,w_1}/D^{2}_{L,\Lambda}$ where $D^{2}_{L,w_0,w_1}$ is the luminosity distance 
 computed (at the considered redshift $z=6$) for considered $(w_0,w_a)$ combination, and $D^{2}_{L,\Lambda}$ is its value in the $\Lambda$CDM 
 case. 
 
We focus on the largest stellar mass bin (centered on $M_*=8\cdot 10^{10}\,M_{\odot}$ assuming a Salpeter IMF) analyzed by  Grazian et al. (2015). 
For each combination $(w_0, w_a)$ we compare the volume-corrected, observed abundance of galaxies $\tilde{\phi}=\phi_{obs}\,f_{Vol}$ with stellar mass $M_*=8\,f_{lum}\,10^{10}\,M_{\odot}$ at $z=6$ with the predicted 
 number density $\phi_{w_0,w_a}(M)$ (eq. 1) of DM halos with DM masses larger than $M=M_*/(F\,f_b)$   for the considered 
 $(w_0, w_a)$ combination. The confidence for the exclusion $P_{excl}$ of each considered DDE model is obtained  
 from the PDF as the probability  that the measured abundance is larger  than number density predicted by the model,  i.e., $P_{exl}(w_0,w_a)=\int_{\phi_{w_0,w_a}}^{\infty} p(\tilde{\phi})\,d\tilde{\phi}$.

\vspace{-0.4cm}
\begin{center}
\scalebox{0.55}[0.55]{\rotatebox{0}{\includegraphics{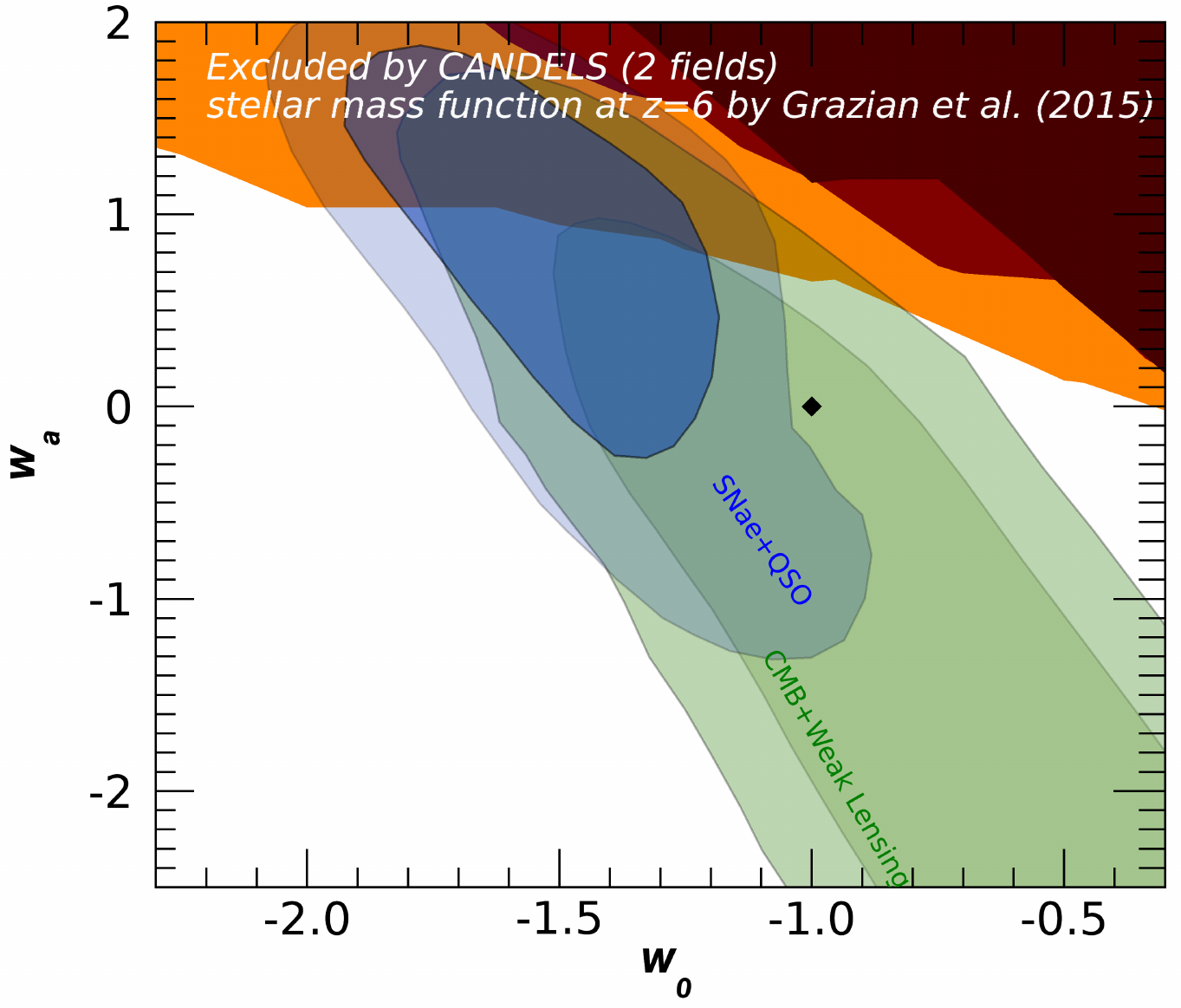}}}
\end{center}
\vspace{-0.4cm }
{\footnotesize Fig. 2. Exclusion regions (2-$\sigma$ confidence level) in the $w_0-w_a$ plane 
derived from the observed CANDELS stellar mass function at $z=6$ (Grazian et al. 2015). The  brown, red, and orange regions correspond to assuming   $F=1$,  $F=0.5$, and $F=0.25$, respectively (see text). Our {\it exclusion} region is compared with the $2-\sigma$ and $3-\sigma$ contours {\it allowed} by 
 CMB+weak lensing (green regions) and by  the combination of the same data with the Hubble diagram of supernovae and quasars 
 (blue region), derived from fig. 4 of Risaliti and Lusso (2019). The black dot corresponds to the $\Lambda$CDM case ($w_0=-1$, $w_a=0$).
}
\vspace{0.2cm }

We show in fig. 2 the region of the $w_0-w_a$ excluded at 2-$\sigma$ confidence level (i.e., $P_{exl}\geq 0.95$) for $F=1$, 
$F=0.5$ and the $F=0.25$ case.  The exclusion region is overplotted to the regions allowed by CMB and weak lensing, and to the region derived 
by the combination of the same data with the Hubble diagram of supernovae and quasars (Risaliti and Lusso 2019). Our probe significantly  restricts the  region in DDE parameter space allowed by other methods. In particular, we 
exclude an appreciable part of the region favored by the  distant quasar method. 

We stress that our method allows for significant improvements when more extended databases will be available in the future. 
To stress this point, we show in  fig. 3 the  constraints that would be obtained from the Grazian et al. (2015) stellar mass function if we decrease by $1/2$ the dispersion in the PDF around the average value. 
This (approximatively) simulates the effects of the larger statistics that would be obtained analyzing the full CANDELS dataset (5 fields).   In fact, the spread $\Delta \phi_{obs}$ in the measured stellar mass function at large masses $M_*\approx  10^{11}\,M_{\odot}$ is dominated by Poisson fluctuations which  account for 70 \% of log $\Delta \phi_{obs}$ 
(while cosmic variance and the uncertainties related to the assumed star formation histories approximatively account for 15\% and 8\%, respectively).
In this case, most of the region allowed by distant quasars would be excluded even in the conservative case $F=0.5$.

\vspace{-0.4cm}
\begin{center}
\scalebox{0.55}[0.55]{\rotatebox{0}{\includegraphics{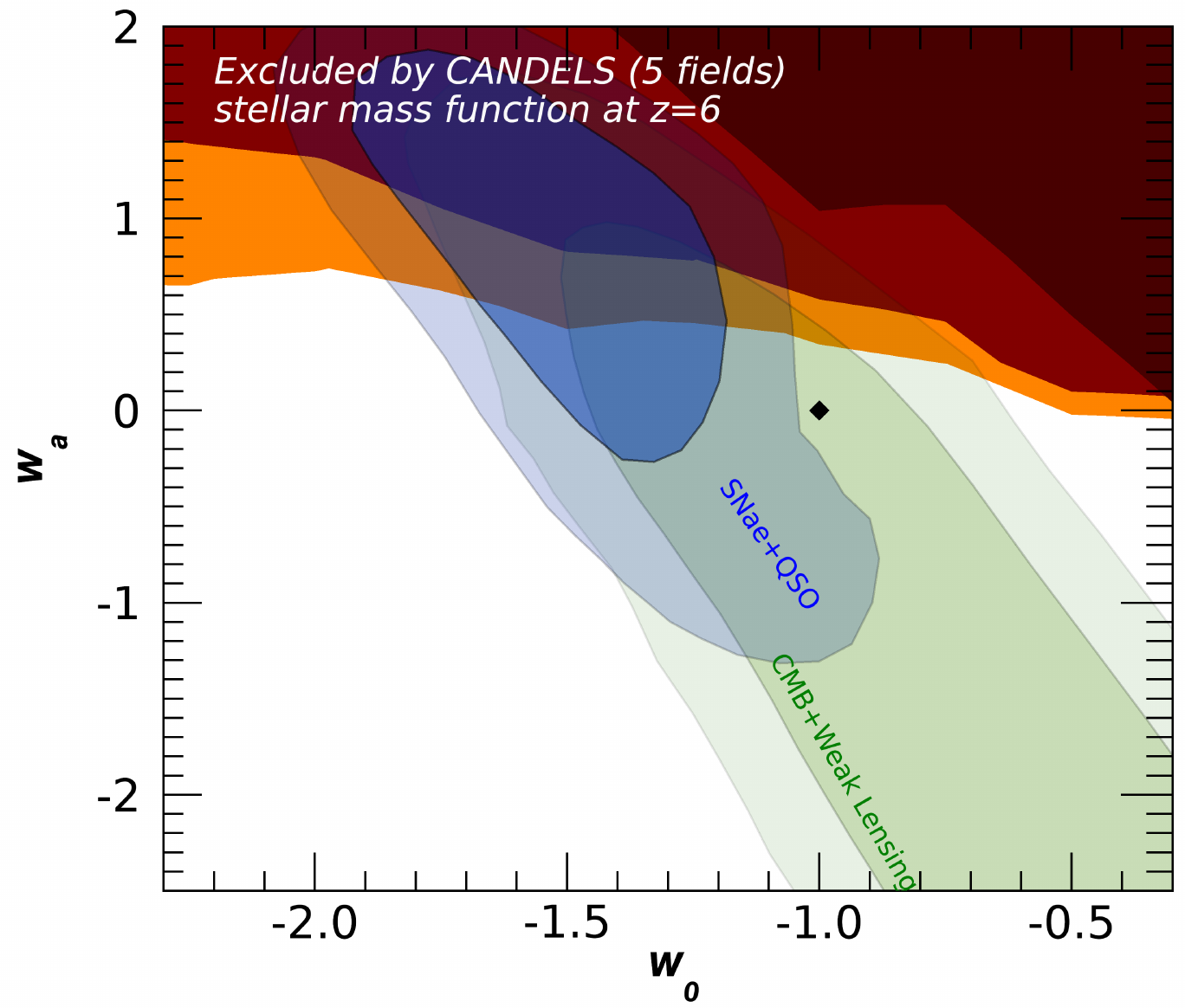}}}
\end{center}
\vspace{-0.4cm }
{\footnotesize Fig. 3. As fig. 2, but assuming the errorbars of the stellar mass function reduced by 1/2, to simulate the inclusion of the full set of CANDELS field. This improvement would exclude most of the region
allowed  by distant quasars.}

Up to this point, we have focused our analysis on the stellar mass function by Grazian et al. (2015). This is because 
it couples large stellar mass coverage at high redshifts (stellar masses extending to $M_*\approx 10^{11}\,M_{\odot}$) with a 
detailed analysis of the uncertainties, including those related to assuming different star formation histories, metallicities, ages, dust extinction,  photometric redshifts, and cosmic variance. Indeed, for this specific measurements we could exploit our previous work  to perform a full computation of the whole PDF through Monte Carlo simulations accounting for the above uncertainties.

Of course, our method can be applied to other estimates of the stellar mass function at high redshifts. Although several observational works exist in the literature (see, e.g., Duncan et al. 2014, Song et al. 2016, Stefanon et al. 2017, Bhatawdekar et al. 2019), to our aim it is essential to compare with observed stellar mass functions that cover high-redshifts ($z\gtrsim 6$) and large masses $M_*\gtrsim 5\cdot  10^{10}$, where the upper limit provided by the halo mass function can be violated in some DDE models 
 (see fig. 1).  E.g., the extremely deep mass functions measured  by Bhatawdekar et al. (2019) for the Frontier Field galaxies do not provide any constraint to DDE models, since at high redshifts ($z\geq 6$) they reach masses $M_*\approx  5\cdot  10^{9}\,M_{\odot}$; these probe DM halo masses below $M\leq 10^{11}M_{\odot}$, 
 where the upper limit provided by the halo mass functions is essentially consistent with  all DDE models (see fig. 1). 
The same argument applies to the measurements by Song et al. (2016) who analyze the CANDELS GOODS-South field to probe the distribution of  stellar masses up to $M_*\lesssim 1.5\cdot 10^{10}M_{\odot}$  at $z=6$ basing on the stellar mass-to-light conversion. 
Stefanon et al. (2017), also basing on stellar mass-to-light conversion, probe somewhat larger stellar masses extending up to $M_*\lesssim 3\cdot10^{10}M_{\odot}$  at $z\approx 6$, but the errorbar  in the most massive bin is too large to allow for effective discrimination among different DDE models. 

A more constraining measurement for our scope has been performed by 
Duncan et al. (2014), who analyzed the CANDELS GOODS-South field. Assuming an observationally-based stellar mass-to-light conversion the above authors measured the stellar mass function  at $z\approx 6$ up to large stellar masses $M_*\approx 10^{11}\,M_{\odot}$. 
Since we cannot compute the full PDF for this observational analysis, the $2-\sigma$ confidence level of exclusion has been derived 
doubling the errorbars presented in Duncan et al. (2014).  To account for the uncertainty related to cosmic variance (not provided by the above authors for the large masses here considered)  we added (in quadrature) the value provided (for the same GOODS-South field) by Song et al. (2016) for different stellar masses at various redshifts. The result is presented in fig. 4, showing constraints on DDE models very similar to the ones derived using the measurement by Grazian et al. (2015). 

\vspace{-0.4cm}
\begin{center}
\scalebox{0.55}[0.55]{\rotatebox{0}{\includegraphics{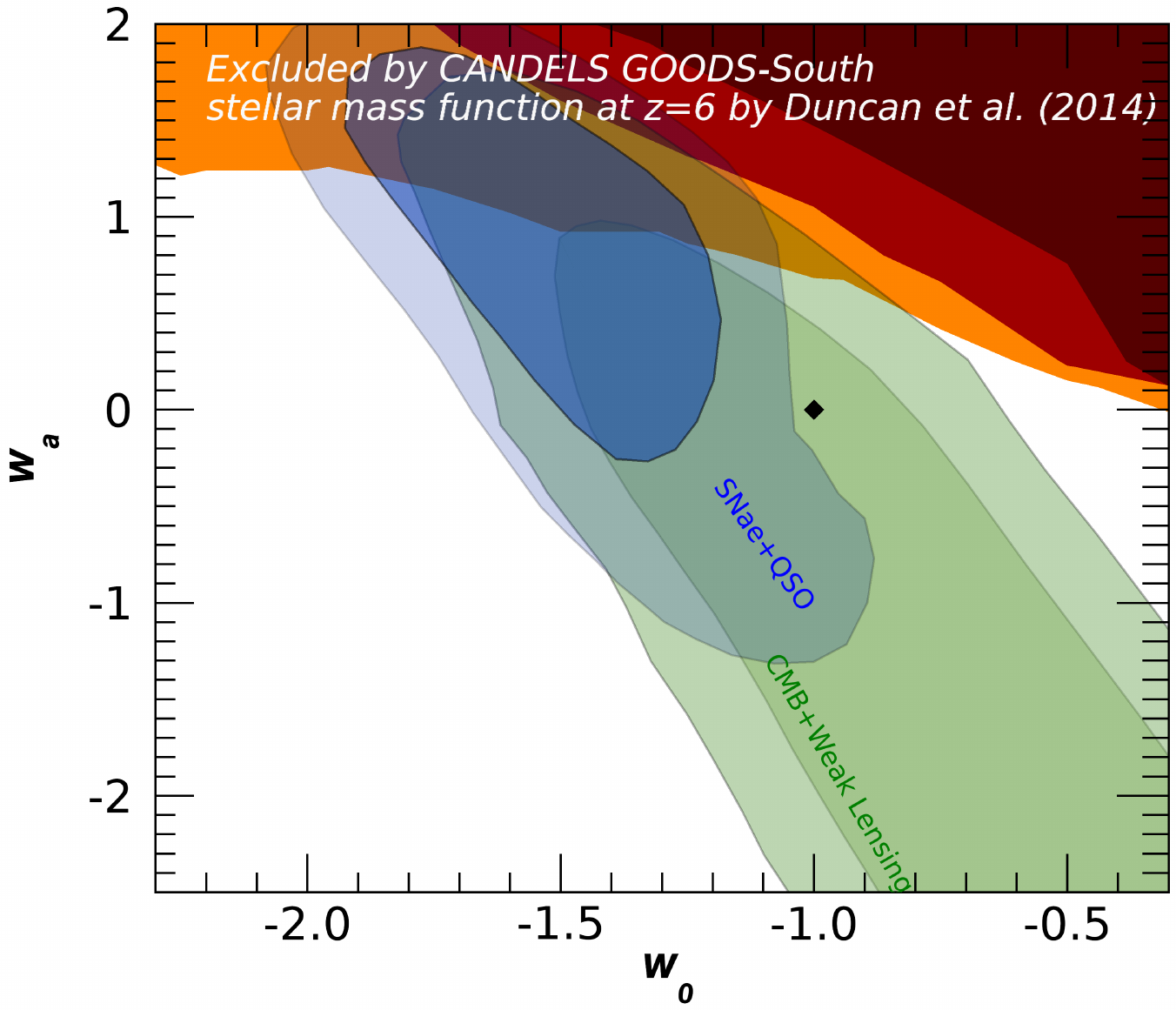}}}
\end{center}
\vspace{-0.4cm }
{\footnotesize Fig. 4. Exclusion regions (2-$\sigma$ confidence level) in the $w_0-w_a$ plane 
derived from the observed CANDELS GOODS-South field stellar mass function at $z=6$ by Duncan et al. 2014). }

Finally, we stress that the constraints presented above are likely to be extremely conservative. Although 
 baryon conversion efficiencies $F$ larger than 0.5 are in principle possible in different DDE models, values closer to 
 the $\Lambda$CDM value $F\approx 0.2$  (derived by abundance matching, see, e.g., by Behroozi 2013, Moster et al. 2018) are much more probable. To address this point in a closer detail,  we have computed  the values of $F$ that would be needed to match  (i.e.,  to lay within the $1-\sigma$ errorbar)  the massive end ($10.75\leq M_*/M_{\odot}\leq 11$) of the observed stellar mass function at $z=4$.  Adopting the stellar mass function at $z=4$ by Grazian et al. (2015), we derive - for each DDE model and for the above range of $M_*$ -  the  baryon efficiency $F$ shown in fig. 5. For all relevant $w_0-w_a$ combinations we obtain $F< 0.5$, while for small values of $w_0\lesssim -1.5$ we obtain $\lesssim 0.25$
   (similar values of $F$ are obtained when comparing with the data by Stefanon et al. 2015; Duncan et al. 2014, Ilbert et al. 2013, Davidzon et al. 2017; even lower values of $F$ are obtained comparing with the data by Song et al. 2016). Notice that when the $\Lambda$CDM model is considered (the black point in fig. 5) our analysis yields a value $F=0.18$, corresponding to a stellar-to-DM mass ratio $M_*/M=F\,f_b=0.025$, in excellent agreement with the value estimated by Behroozi et al. (2013; see also fig. 9 in Song et al. 2016)  from the abundance matching in the $\Lambda$CDM case. 

\vspace{-0.4cm}
\begin{center}
\scalebox{0.5}[0.5]{\rotatebox{0}{\includegraphics{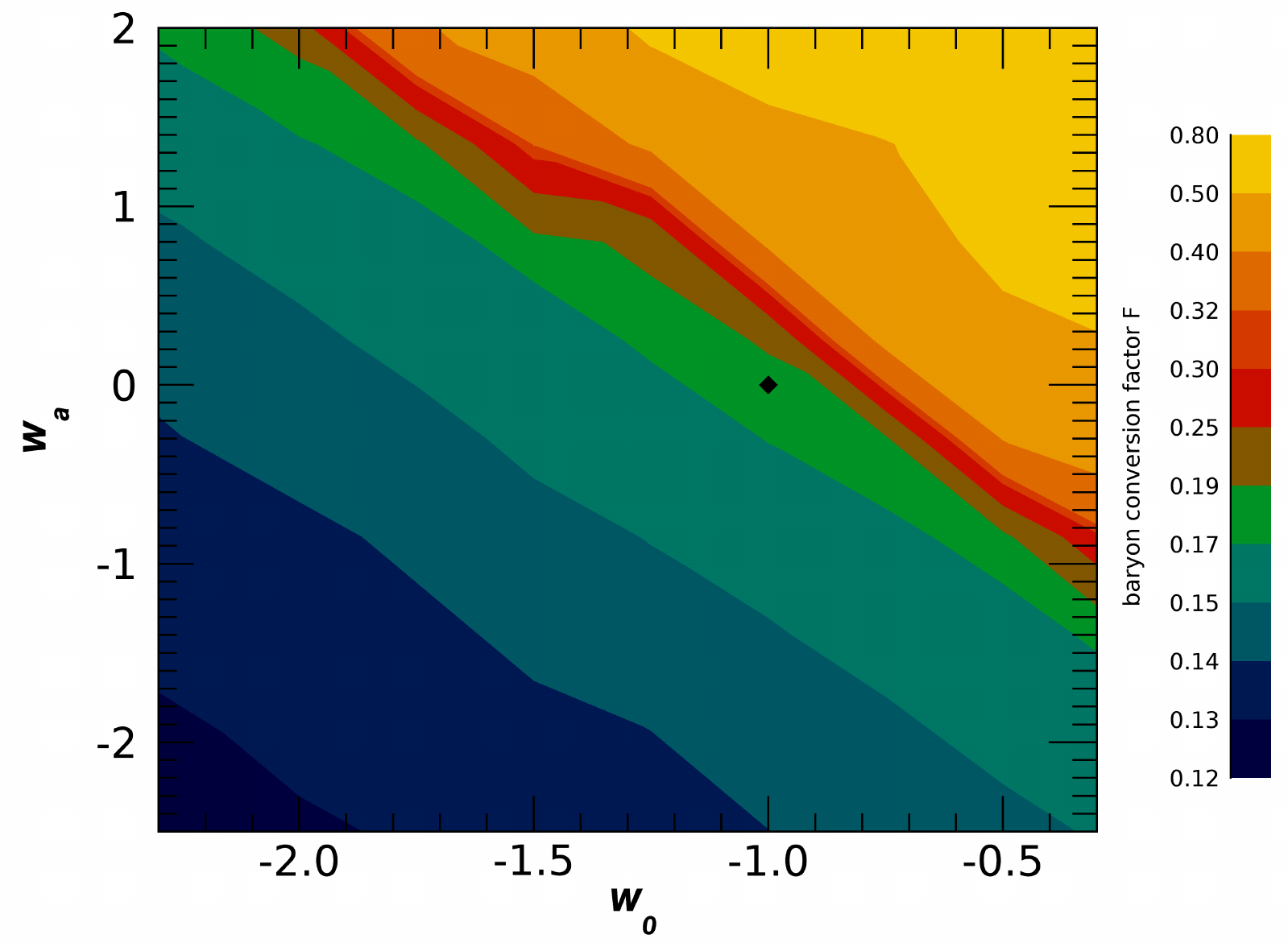}}}
\end{center}
\vspace{-0.4cm }
{\footnotesize Fig. 5. For each combination $w_0-w_a$, we show the value of the baryon efficiency $F$ (color coded as 
shown in the vertical bar) needed to match the 
stellar mass function at $z=4$ for $10.5\leq log(M_*/M_{\odot})\leq 11$}

\vspace{-0.4cm}
\begin{center}
\scalebox{0.55}[0.55]{\rotatebox{0}{\includegraphics{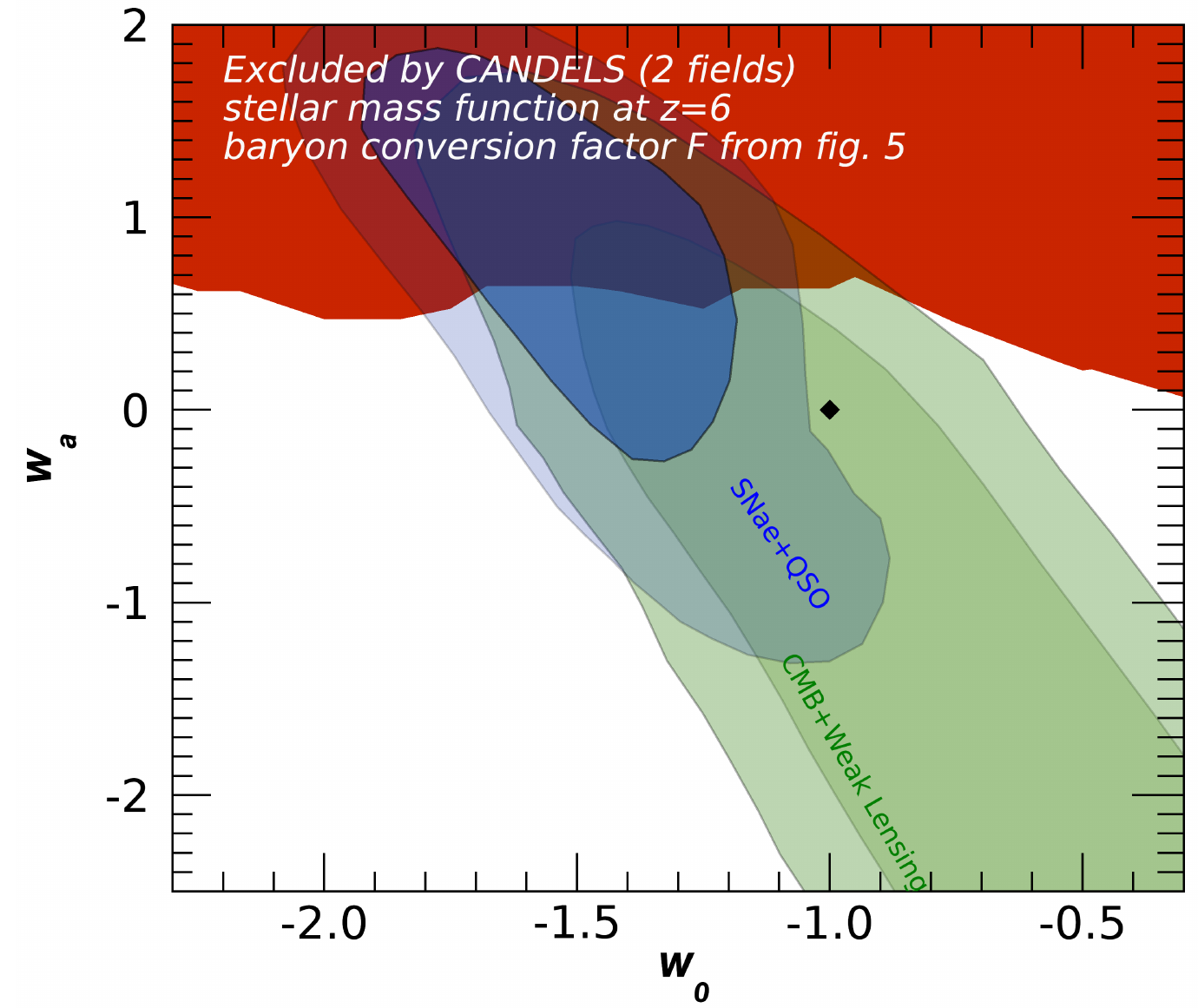}}}
\end{center}
\vspace{-0.4cm }
{\footnotesize Fig. 6. As in fig. 2, but assuming the baryon efficiency $F$ from fig. 5.}

Under the assumption of constant $F$ between redshift $z=4$ and $z=6$, we can then use the values of $F$ shown in fig. 5 to recompute the constraints on DDE models from the comparison with the CANDELS  stellar mass function at $z=6$. 
The result, shown in fig. 6, shows that in this case we obtain even stronger constraints, since the values of $F$ for the different 
DDE models are now within the range $0.2\lesssim F\lesssim 0.25$ for a wider set of  combinations $w_0-w_a$. 
This shows that - if the baryon fraction $F$ has a slow evolution in DDE models between $z=4$ and $z=6$ -  the assumptions used to derive the constraints  in fig. 2-4 are indeed conservative. 

\subsection{Massive galaxies detected in sub-mm at  $z\approx 4$} 
The above population of galaxies (identified in rest-frame optical and ultraviolet) is known to under-represent the most massive galaxies, which have rich dust content and/or old stellar populations. These are however detectable at submillimetre wavelengths . 
Recently, Wang et al (2019) performed detailed submillimeter (870 micrometres) observations at the Atacama Large Millimeter Array (ALMA) of 
a sample of Spitzer/Infrared Array Camera (IRAC)-bright galaxies. They detected 39 
star-forming objects at $z > 3$, which are unseen in even the deepest near-infrared (H-band) imaging with the Hubble Space Telescope (H-dropouts), which proved to be massive galaxies with  stellar mass extending up to $M_*\approx 3\cdot 10^{11}\,M_{\odot}$, with a median mass $M_*\approx 4\cdot 10^{10}\,M_{\odot}$. 

For such objects we follow a procedure similar to what explained in the previous Section. We compute the number density of galaxies 
 with stellar mass in the bin  $10.25\leq$ log($M_*/M_{\odot})\leq 10.75$ (dominating the statistics of observed objects) at redshifts $z=4.5-5.5$ , and derive the corresponding 2-$\sigma$ lower limit $\phi_{low}(M_*)=1.8\cdot 10^{-5}$ Mpc$^{-3}$. To relate the observed stellar mass $M_*$ to the DM mass $M$ we adopt the highly conservative assumption $M=M_*/f_b$. We then computed the number density $\phi_{w_0,w_a}$ of DM halo mass corresponding to 
 the observed $M_*$ for different combinations ($w_0$, $w_a$), and compare it with the observed $2-\sigma$ lower limit $\phi_{low}$. 
For each combination ($w_0$, $w_a$), observed number densities and stellar masses (measured assuming a $\Lambda$CDM cosmology) have been  
  rescaled with the factors $f_{Vol}$ and $f_{lum}$ (see Sect. 2.1). 
 The comparison allows us to exclude (at 2-$\sigma$ confidence level) the combinations  ($w_0$, $w_a$) for which $\phi_{w_0,w_a}<\phi_{low}$.  The results is show as a brown exclusion region in fig. 7.

Of course, the above approach is very conservative, since we assumed 
that the whole baryonic mass is in stars, and that the baryon mass of DM haloes is related to the DM mass through the universal baryon fraction 
(no loss of baryons). In fact, the very fact that the objects are characterized by a high star formation rate $\gtrsim 200\,M_{*}$/yr indicates that 
a sizable fraction of baryon is in the form of gas. Properly accounting for such gas fraction would yield larger values $M$ associated to the 
 observed $M_*$ and - hence - tighter constraints. Although we have attempted to estimate the gas mass for the ALMA-detected H-dropout galaxies from the sub-mm continuum by converting the dust mass through the dust/gas ratio, the inferred gas masses are affected by large uncertainties (they span a range between $5\cdot 10^{9}\,M_{\odot}$ and $5\cdot 10^{10}\,M_{\odot}$), related to photometric redshifts (uncertainties are particularly critical given the steep shape of the spectrum in the far-IR), the adoption of a single and simplified gray body at the average temperature, and the adoption of the mass-metallicity at z=3.5 for all sources.
 
\vspace{-0.2cm}
\begin{center}
\scalebox{0.55}[0.55]{\rotatebox{0}{\includegraphics{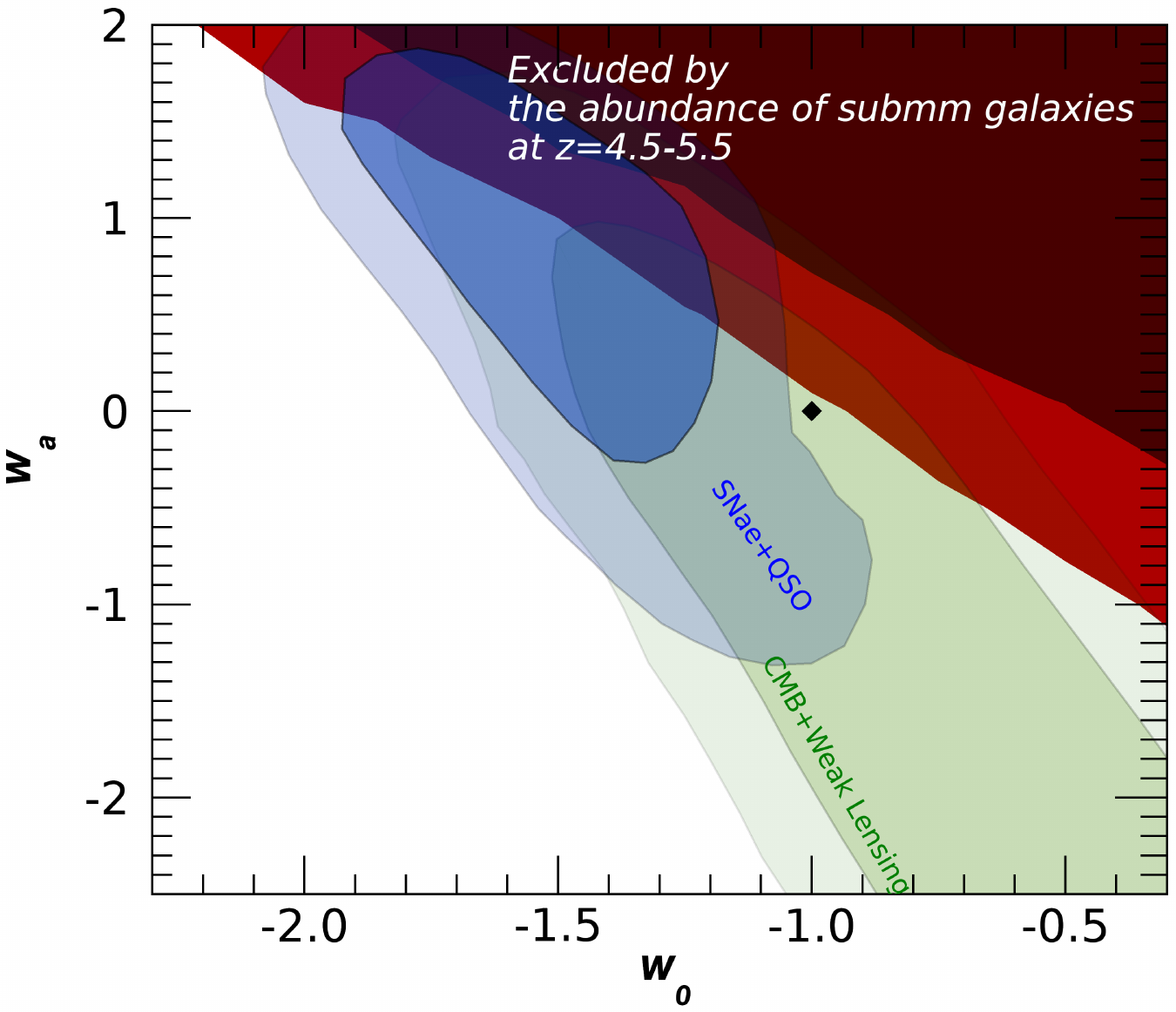}}}
\end{center}
\vspace{-0.4cm }
{\footnotesize Fig. 7. Exclusion regions (2-$\sigma$  confidence level) in the $w_0-w_a$ plane (see text) 
derived from the observed abundance $\phi_{obs}$ of luminous submm galaxies  at $z = 4.5 - 5.5$ (Wang et al. 2019). 
The brown region corresponds to assuming  the observed stellar masses $M_*=M/f_b$ to be  related to the DM mass through the 
baryon fraction $f_b$. The red regions corresponds to adopting the DM mass derived from the measured cross-correlation function of 
H-dropouts (see text). 
}

To bypass the uncertainty related to the gas fraction, and to derive more realistic constraints 
 for DDE models, we analyzed the clustering properties of the H-dropouts. We base on the procedure adopted by Wang et al. (2019)  who estimated the two-point angular cross correlation function $\omega(\theta)$ of H-dropouts with all CANDELS galaxies in the redshift range $3.5\leq z\leq 5.5$. Assuming a power-law form for the cross correlation $\omega(\theta)=A_{\omega}\,\theta^{-\beta}-IC$ (with  $\beta=0.8$ and IC  the integral constraint in eq. 4 of Wang et al. 2019), the above authors derived the amplitude  $A_{\omega}$. This was related to the correlation length $r_0$ by the Limber equation (Croom and Shanks 1999; He, Akiyama, Bosch et al. 2018)
\begin{equation}
r_0=\Bigg[ A_{\omega}\,{c\over H_0\,Q } {\int N_H(z)dz\int N_G(z)dz\over \int N_H(z)\,N_G(z)\chi^{1-\gamma}(z)\,E(z)dz}  \Bigg ]^{1/\gamma}
\end{equation}
where $\gamma=\beta+1$, the constant $Q=\Gamma(1/2)\,\Gamma(\gamma/2-1/2)/\Gamma(\gamma/s)$ is a combination of $\Gamma$ functions, 
$ \chi (z)$ and $E(z)$  are given in eq. 5 and 6, and $N_H(z)$ and $N_G(z)$ are the redshift distributions of H-dropouts and CANDELS galaxies. 
The correlation length $r_0$ was then converted to galaxy bias through the relation (Peebles 1993) 
\begin{equation}
 b={72\over (3-\gamma)(4-\gamma)(6-\gamma)2^{\gamma}\,\sigma_8(z)}\,\Big[{r_0\over 8\,h^{-1}{\rm Mpc}}\Big]^{\gamma}˜,
 \end{equation}
that we assume to hold in all DDE models. Here  $\sigma_8(z)$ is the amplitude of the dark matter fluctuation on the scale of 8 $h^{-1}$ Mpc. 
The DM mass is then derived from the relation $b=1+\big[\nu(M,z)-1\big]/\delta_c$ (Mo and White 2002). For the standard $\Lambda$CDM case the above procedure yields $M=10^{13\pm 0.3}\,M_{\odot}$ for the average DM mass (Weng et al. 2019). 

For our comparison with DDE predictions, we cannot take the above DM mass at face value, since it has been derived assuming a $\Lambda$CDM 
cosmology. In fact,  
 for generic DDE cosmologies, the above value will (weakly) change due to two factors: i) the Limber equation (eq. 10) relating the observed $A_{\omega}$ to $r_0$  depends on cosmology through the functions $E(z)$ and $\chi(z)$ (Sect. 2) ; and ii) the different growth factor  (Sect. 2, eq. 7 and below) affects the quantities $\nu(M,z)$ and $\sigma_8(z)$ entering the computation of the  average mass $M$ (eq. 11 and below). Thus, we computed the maximal effect of cosmology on the value of $M$ derived by Wang et al. (2019) when  our grid of values for the of combinations ($w_0,w_a$) is considered. Assuming the same measured angular cross correlation amplitude $A_{\omega}$, we considered the effect of 
 different cosmologies on the derived 3D correlation length (eq. 10) and on the bias factor (eq. 11).  
 We found that $M=10^{13}\,M_{\odot}$ constitutes a (2-$\sigma$) lower limit for the value of the DM mass derived from cross correlation for any DDE model we considered. We then conservatively computed the DDE number density of objects with such a DM mass and compared it with 
 the observed number density of H-dropouts with stellar mass $M_*=10^{10.5}\,M_{\odot}$ (the average stellar mass of the sample). The resulting exclusion region in the ($w_0,w_a$) plane is shown in red in fig. 7.

\subsection{SPT0311–58 at $z = 6.9$} 

The most massive  system detected at $z\geq 6$ is a far-infrared-luminous object at redshift $z=6.9$ originally identified in the 2500 deg$^{2}$ South Pole Telescope (SPT) survey  (Marrone et al. 2018). Observation in the optical with the HST, infrared observations with the Spitzer Space Telescope, Gemini Optical/IR imaging and spectroscopy subsequently allowed for a characterization of this source. 
High-resolution imaging revealed this source (denominated SPT0311–58) to be a pair of extremely massive star-forming galaxies, with the larger  galaxy (SPT0311–58W) forming stars at a rate of 2900 $M_{\odot}$/yr. 
An elongated faint object seen at optical and near-infrared wavelengths is consistent with a nearly edge-on spiral galaxy at $z\approx  1.4$ acting as a gravitational lens for the source, with an estimated magnification $\mu=2$. 

Measurements of the far-infrared continuum with the Atacama Large Millimeter/submillimeter Array (ALMA) led to estimate a huge 
H$_2$ gas mass 
ranging from $M_{H_2}\approx (7.6\pm 2)\cdot 10^{10}\,M_{\odot}$ (based on CO luminosity converted with a
 standard value $\alpha_{CO}= 1$ km s$^{-1}$ pc$^{2}$) to 
$M_{H_2}\approx (3.1\pm 1.9)\cdot 10^{11}\,M_{\odot}$ (estimated from a radiative transfer model in Strandet et al. 2017). 

In the following we shall adopt the latter value as a baseline, since it is  based  on a detailed fit with a radiative transfer model, built ad-hoc to study the ISM properties of this very same object. As explained in detail in Strandet et al. (2017 and references therein), the model provides a fit to the FIR continuum sampled in 7 broad bands, three CO lines and the [CI] line observed in this specific object by ALMA, Herschel, SPT, APEX and ATCA.  As mentioned in Marrone et al. (2019) such estimate is more accurate than the one obtained by converting the CO line. In fact the latter is derived using a standard {\it average} ULIRG-like $\alpha_{CO}=1$ 
km s$^{-1}$ pc$^{2}$. However, such a quantity 
strongly depends in specific properties of the considered object like star formation, and metallicity (see Bolatto et al. 2013), 
so that a large scatter around the average value in different objects is expected (as in fact observed, see, e.g., Weiß
et al. 2007, Papadopoulos et al. 2012). 
Indeed,  the radiative transfer model applied to SPT0311–58 yields $\alpha_{CO}\approx 4$ km s$^{-1}$ pc$^{2}$. The authors 
 explain the difference with respect to the typical, ULIRG-like factor by the much higher ISM density in this object.

To estimate the DM mass associated to such object, we cannot follow the procedure adopted in Marrone et al. (2018), since they derive the gas-to-DM conversion factor from abundance matching techniques (see, e.g., Behroozi et al. 2018) which cannot be safely considered as a baseline for generic DDE models, since they base on the $\Lambda$CDM halo mass function. 
 
Thus, to estimate a conversion fraction from the observed $H_2$ mass to the DM mass we first adopt 
the conservative assumption that the total baryonic mass $M_b=M_*+M_{gas}$ (here $M_{gas}$ is the total gas mass) is related to the DM mass through the baryon fraction $M=(M_*+M_{gas})/f_b$. Although  no stellar light is convincingly seen from SPT0311–58W (probably due to the large extinction)  a lower limit on the stellar content can be inferred from existing measurements of the molecular gas fration $f_{H_2}=M_{H_2}/(M_*+M_{H_2})$. Measurements of high-z star forming galaxies (ranging from relatively quiescent BzK galaxies to dusty starbursts), suggest  $f_{H_2}=0.2-0.8 $ (e.g. Daddi et al., 2010; Tacconi et al., 2010; Geach et al., 2011; Magdis et al., 2012; Combes et al., 2013; Tacconi et al., 2013, see Casey, Narayanan, Cooray 2014 for a review). However, all theoretical models 
(Benson et al. 2012; Lagos et al. 2012; Fu et al. 2012; Popping et al. 2014;  Davé et al. 2012; see also Gabor and Bournaud 2013; Ginolfi et al. 2019) predict typically smaller values in the range $f_{H_2}\lesssim 0.5$. One possible solution to this mismatch has been offered by Narayanan et al. (2012), who suggested that the canonical conversion factor $CO$-$H_2$  was too large for the most extreme systems at high-redshift, and that the correct observed gas fractions are in the 
range $f_{H_2}=0.1-0.4$.  Similar conclusions are drawn by Tacconi et al. (2013) who suggested that the tension between galaxy gas fractions measured in observations and simulated galaxies may owe to  incomplete sampling of galaxies.

Even assuming that $H_2$ constitute $80\%$ of the gas mass (i.e., $f_g\equiv M_{H_2}/M_{gas}=0.8$) at high redshifts (an upper limit according to Lagos et al. 2011, 2014) the estimated baryonic mass $M_b=M_{gas}+M_*=M_{H_2}(f_{H_2}+f_{g}- f_{H_2}f_{g})/
f_{H_2}f_{g}$ takes the value $M_b=1.4\,M_{H_2}$ if we adopt the most conservative estimate $f_{H_2}=0.8$, and $M_b=2.75\,M_{H_2}$  if we adopt the estimate $f_{H_2}=0.4$ suggested by theoretical models and by the effects suggested by Narayanan et al. (2012) or by  Tacconi et al. (2013). 
Although the latter works refer to galaxies at $z\leq 4$, considering the above range of uncertainty represents the best we can do 
with present data and the available theoretical predictions.  
This leads to associate to the observed $M_{H_2}$ a DM mass $\overline{M}=M_b/f_b=2\cdot 10^{12}\,M_{\odot}$ in the most conservative case, and to  $\overline{M}=M_b/f_b\approx 6\cdot 10^{12}\,M_{\odot}$ in the other case; we will consider both values in the following analysis.  Even larger DM mass would correspond to the observations if the object lost  the majority of
its  molecular gas content.

To estimate the rareness of such a system in all the considered DDE cosmologies, 
we compute the Poisson probability of finding such a massive object within the  volume probed by the SPT survey, for different  combinations $(w_0, w_a)$. Following the method in Harrison and Hotchkiss (2013) as done in Marrone et al. (2018) for the $\Lambda$CDM cosmology, we first compute from eq. (3) the number $N(M,z)$ of systems with mass $M$ and higher at redshift $z$ and higher expected in the sky area $f_{sky}$ covered by the SPT survey, for a grid of values of $M$ and $z$. Then we compute such a number $N(\overline{M},\overline{z})$ for the  values $\overline{M}$ and $\overline{z}$ associated to the observed systems (i.e., $\overline{z}=6.9$ and $\overline{M}=2-6\cdot 10^{12}\,M_{\odot}$ as discussed above). Finally, we consider the number $N_{rare}$ defined as $N(M,z)$ 
computed only for the masses $M$ and redshifts $z$ for which $N(M,z)\geq  N(\overline{M},\overline{z})$, as discussed in Harrison and Hotchkiss (2013). The Poisson probability of observing at least one system with both greater mass and redshift than the one which has been observed is 
\begin{equation}
R_{>\overline{M},>\overline{z}}=1-exp(-N_{rare})
\end{equation}
The above probability depends on the region of the $M-z$ plane to which the SPT survey is sensitive (which provides the lower limit 
for the integration over redshift and mass in eq. 3), and on $f_{sky}$. Following Marrone et al. (2018) we assume that the survey is complete for $z\geq 1.5$ and for $M\geq 10^{11}\,M_{\odot}$, a conservative assumption as discussed in detail by the above authors.
To take into account the uncertainties in the measured value of  $M_{H_2}$ (in turn affecting the corresponding DM mass $M$) we followed Harrison \& Hotchkiss (2013) and convolved  eq. 12 with the probability of measuring a given $M_{H_2}$, assuming a Gaussian distribution around the central value $M_{H_2}=3.1\cdot 10^{11}\,M_{\odot}$ with standard deviation $1.9\cdot 10^{11}\,M_{\odot}$. As for the 
 effective fraction of the sky  $f_{sky}=\Omega_{sky}/(41253$ deg$^{2}$) entering eq. 3, the total area corresponding to the SPT survey is $\Omega_{sky}=2500$ deg$^{2}$.  
 However, Marrone et al. (2018) noticed that the effective survey area is potentially much smaller. In fact, most of the objects in the survey are strongly lensed, indicating that a source must be gravitationally lensed to exceed the  20mJy threshold for inclusion in redshift follow up observations. Given the uncertainties related to properly accounting for such an affect, we show our results for both the total 
  area ($\Omega_{sky}=2500$ deg$^{2}$) and for an effective area reduced by 1/10  ($\Omega_{sky}=250$ deg$^{2}$) to illustrate the effect of such an uncertainty  (Marrone et al. considered an even more extreme case $\Omega_{sky}=25$ deg$^{2}$). 

For  each combination $(w_0, w_a)$, we compute the expected number of systems like SPT031158 detectable in the SPT survey. 
 Then we associate a rareness to the resulting predicted number after eq. 10, and we compute the associated exclusion regions in 
 the $w_0-w_a$ plane. The result (2-$\sigma$ confidence level) is shown in fig. 8  for the case $\Omega_{sky}=2500$ deg$^{2}$, for the two considered values $\overline{M}=2\cdot 10^{11}\,M_{\odot}$ (red region) and $\overline{M}=6\cdot 10^{11}\,M_{\odot}$ (orange region). 
We also show as a dashed line the bound of the exclusion region that would be obtained for 
$\overline{M}=0.5\cdot 10^{11}\,M_{\odot}$. This would correspond to a  gas mass 
$M_{H_2}= (7.6\pm 1.9)\cdot 10^{10}\,M_{\odot}$ based on CO luminosity converted with a
 standard value $\alpha_{CO}= 1$ km s$^{-1}$ pc$^{2}$ with $f_{H_2}=0.8$.
  
\vspace{-0.4cm}
\begin{center}
\scalebox{0.55}[0.55]{\rotatebox{0}{\includegraphics{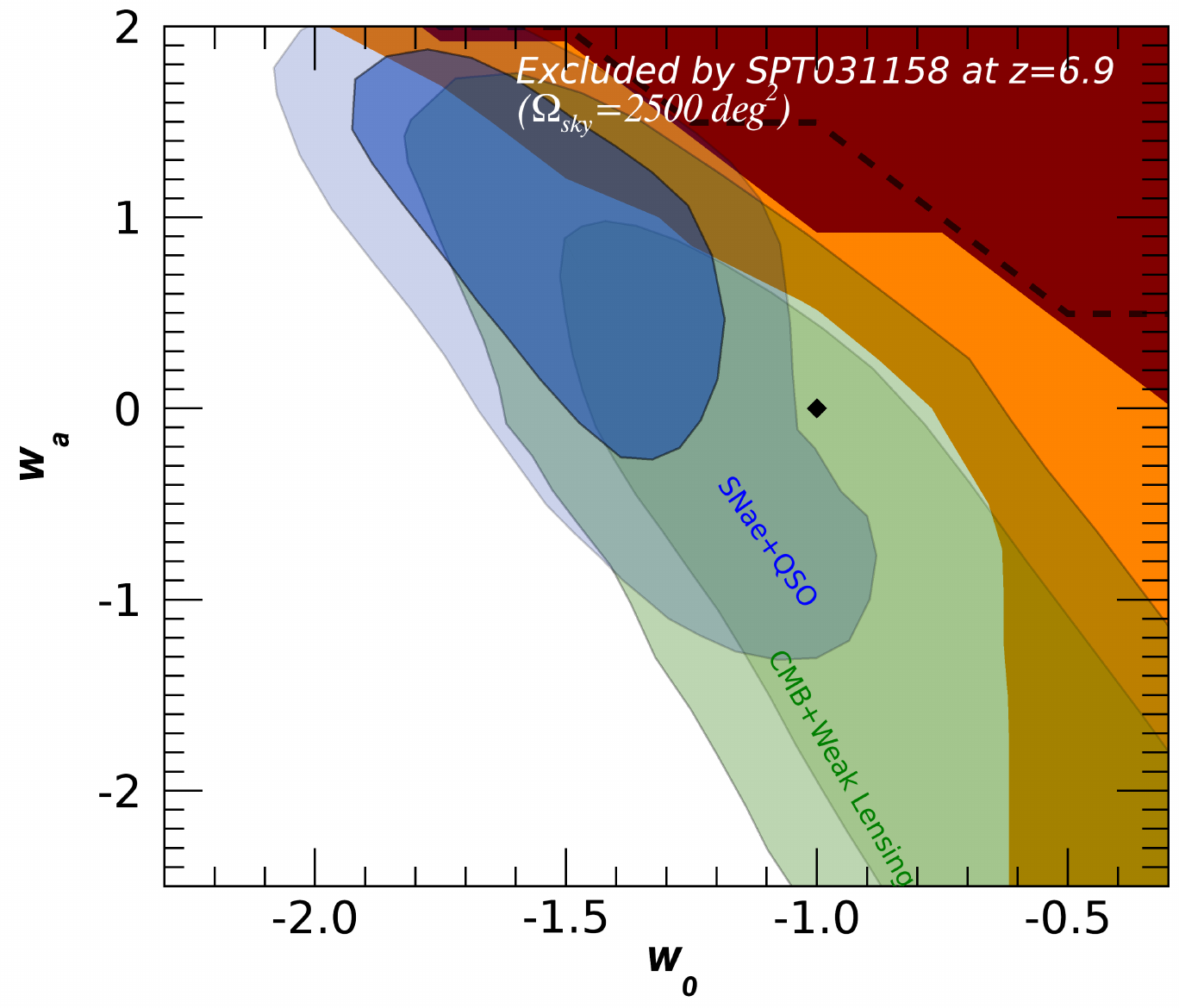}}}
\end{center}
\vspace{-0.4cm }
{\footnotesize Fig. 8. Exclusion regions (2-$\sigma$ confidence level) in the $w_0-w_a$ plane (see text) 
for two different inferred DM mass of SPT0311–58: $2\cdot 10^{12}\,M_{\odot}$ (red area) and 
 $6\cdot 10^{12}\,M_{\odot}$ (yellow area). In both cases the full SPT survey area  $\Omega_{sky}=2500$ deg$^{2}$ has been assumed. The dashed line represents the bound of the exclusion region that would be obtained for 
$\overline{M}=0.5\cdot 10^{11}\,M_{\odot}$, corresponding 
 to the gas mass derived from the CO  measurement with conversion factor  $\alpha_{CO}= 1$ km s$^{-1}$ pc$^{2}$, see text.
}
\vspace{0.2cm }

In the case $\overline{M}=6\cdot 10^{11}\,M_{\odot}$, corresponding to assuming the value $f_{H_2}=0.4$ for the $H_2$ gas fraction, a major portion of the 
  $w_0-w_a$ is excluded, although the $\Lambda$CDM case ($w_0=-1$, $w_a=0$) remains allowed. The excluded region includes both the   larger $w_a$ cases allowed by the quasar method (blue region) and the  cases $w_0\geq -0.6$ allowed by the CMB+ weak lensing results, showing the potential impact of our results.  Even tighter constraints are obtained for the case $\Omega_{sky}=250$ deg$^{2}$ shown in fig. 9.

\vspace{-0.4cm}
\begin{center}
\scalebox{0.55}[0.55]{\rotatebox{0}{\includegraphics{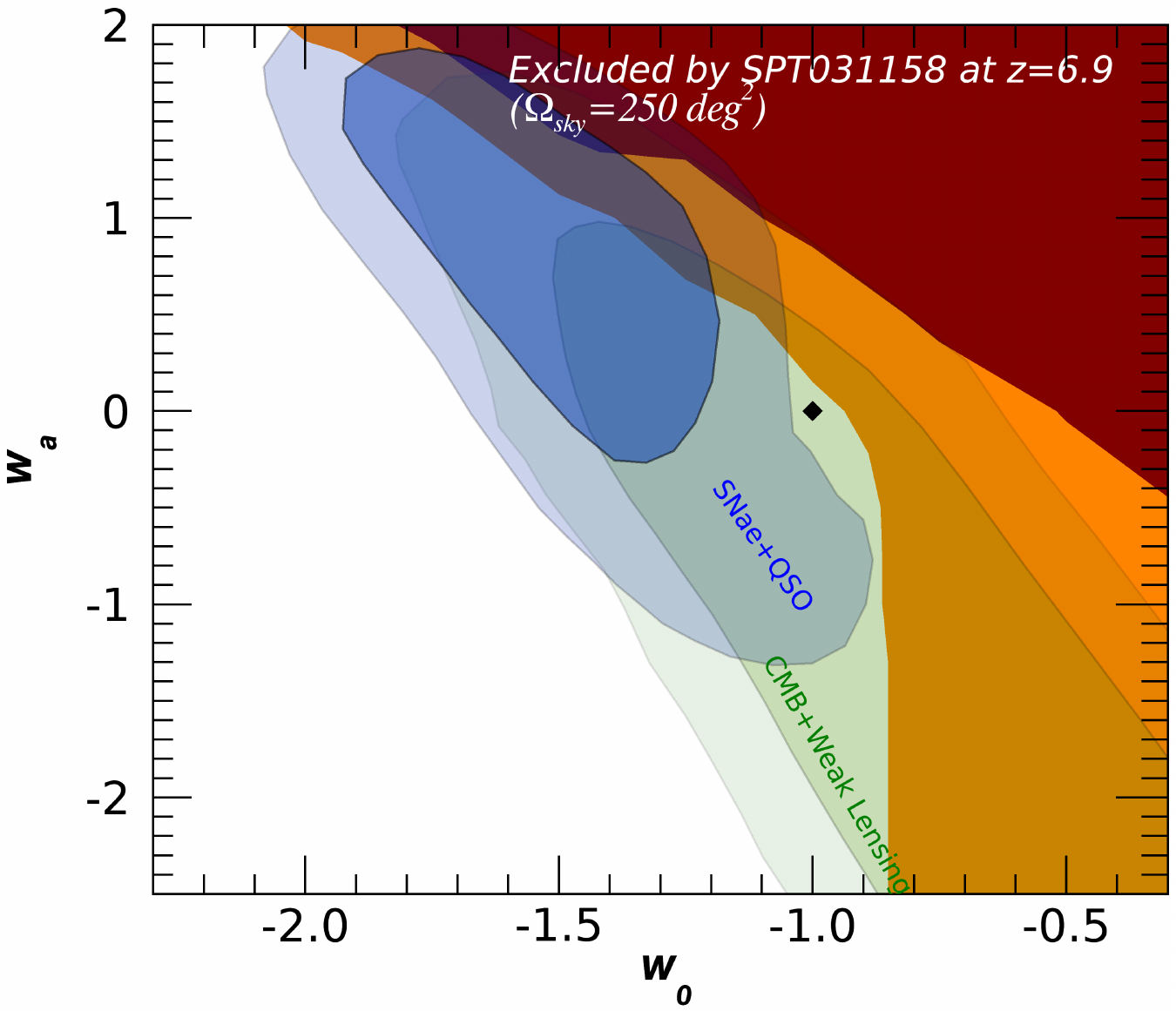}}}
\end{center}
\vspace{-0.4cm }
{\footnotesize Fig. 9. Same as fig. 8, but assuming an affective SPT area $\Omega_{sky}=250$ deg$^{2}$.
}

\subsection{Combining the different probes} 

In the previous sections (3.1-3.3) we have shown the potentiality  of different observables as constraints on DDE models, and discussed how the effectiveness of 
each  probe relies on how much the observed baryon-to-DM mass ratio is suppressed with respect to the 
baryon fraction limit. While future observations will allow for a more precise determination of the gas and stellar mass fractions  (see discussion in Sect.4 below), strong constraints  can be derived - even under the most conservative assumptions - combining all the  probes presented in Sect. 31.-3.3. 
 In fact, the probabilities for each combination ($w_0$,$w_a$) to be consistent with  each of the considered observations are independent. Thus we can derive a combined constraint by multiplying the probabilities of being consistent with each probe. The resulting exclusion region is shown in fig. 10 adopting - for each probe in Sect. 3.1, 3.2 and 3.3 - the most conservative assumption for the relation between the observed baryonic component and the DM mass $M$: 
  For the comparison with the CANDELS field we assume that the observed stellar mass is $M_*=0.5\,f_b\,M$ (i.e., $F=0.5$, see Sect. 3.1);  
  For the comparison with the abundance of submm galaxies (Sect. 3.1) we assume that the observed stellar mass  is related to $M$ by the baryonic  fraction limit; As for the rareness of SPT031158, we take the conservative values for the gas mass fraction leading to 
  a DM mass estimate $M=2\cdot 10^{12}\,M_{\odot}$ (see Sect. 3.3), and we consider the whole survey area ($\Omega_{sky}=2500$ deg$^{2}$). 
  
Inspection of fig. 10 shows that a major fraction of the parameter space favored by distant quasars combined with CMB and weak lensing is excluded at   2-$\sigma$ confidence level, {\it independently on the details of the assumed baryon physics}. 
   
\vspace{-0cm}
\begin{center}
\scalebox{0.55}[0.55]{\rotatebox{0}{\includegraphics{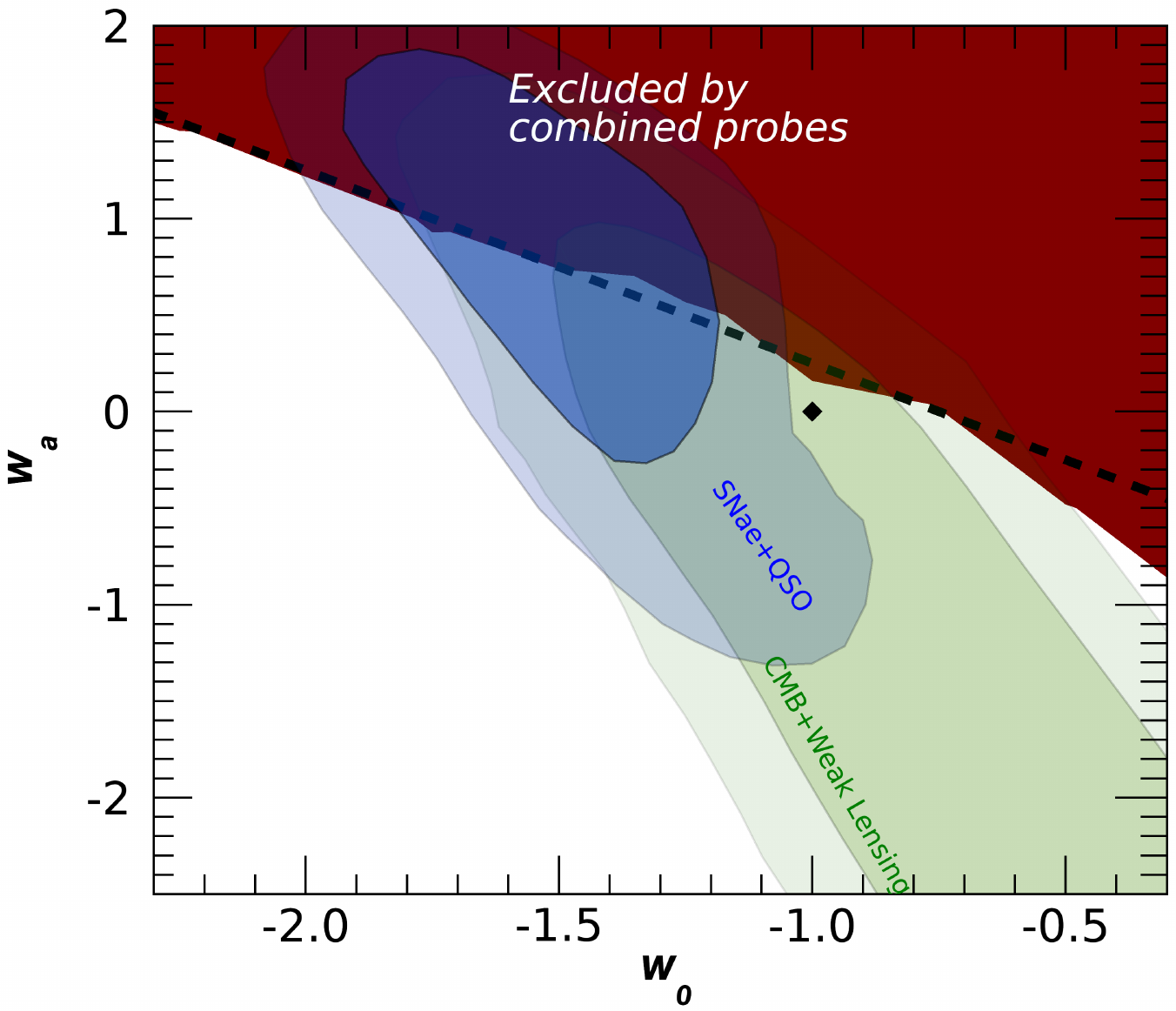}}}
\end{center}
\vspace{-0.4cm }
{\footnotesize Fig. 10. Exclusion regions (2-$\sigma$ confidence level)  in the $w_0-w_a$ plane 
derived from combining the different probes. For each observable, the most conservative case has been considered: 
for the CANDELS field we have assumed $F=0.5$, for SPT031158 we have taken a DM mass $M=2\cdot 10^{12}\,M_{\odot}$, and 
for submm galaxies we have converted stellar masses to DM mass assuming $M=M_*/f_b$.  
The dashed line shows the analytical approximation for the boundary of the excluded region $w_a= -3/4-(w_0+3/2)$.
}

\section{Conclusions and Discussion}

We have computed the abundance of massive systems
predicted in different dynamical dark energy
(DDE) models at high redshifts $z \approx 4-7$. Such predictions
have been compared with different observational
probes: the bright end of the stellar mass function at
$z \geq 6$, the space density of luminous submm galaxies
at $z = 4 - 5$, and the rareness of the extreme hyperluminous
infrared galaxy SPT031158 at $z \approx 7$. \\
We have derived exclusion regions in the parameter space
$ w_0 - w_a$ of DDE models from each of the above probes.
Adopting the most conservative assumptions for the
ratio between the observed baryonic component and
the DM mass, we have combined the above results
to derive conservative, robust constraints for the parameter
space of DDE models, that do not depend on
the details of the baryon physics involved in galaxy
formation.  In addition our results do not depend on the 
nature of the DM component, when present limits on the mass
of DM particle candidates $m_X\gtrsim 3$ keV
(see, e.g., Viel et al. 2013; Menci et al. 2016) are taken into account. In fact, 
 for DM particle masses in the keV range (Warm Dark Matter) 
the associated power spectrum (Bode, Ostriker, Turok 2001;  Destri, de Vega, Sanchez 2013)
on the mass scales investigated in this work $M\geq 10^{10}\,M_{\odot}$ 
is identical to the CDM form assumed here, and our results are unchanged.

\subsection{Implications of our Results}

\begin{itemize}
\item{ 
When the most conservative values concerning the
baryon-to-DM mass are assumed, our combined results
allow to rule out DDE models with  
$$
w_a \geq -3/4 -(w_0 + 3/2)
$$
as displayed in fig.10, thus \textit{excluding a major fraction}
of the parameter space favored by the quasar distances
(Risaliti and Lusso 2019), including the best-fit combination
$w_0 \approx - 0.8$ and $w_a = -1.5$ obtained with such a
probe.}

\item{ 
Our results leave open the possibility 
that the
present tension in the value of $H_0$ between the values
derived from Planck and those obtained from local
luminosity distance measurements be solved in DDE
models, since combinations $(w_0, w_a)$ that allow
to reconcile the different observations include values
outside our exclusion region (see Di Valentino
et al. 2017; Zhao et al. 2017). 
 }
\item{ On the other hand, our results {\it almost entirely
rule out} the quintessence models where initially
$w > -1$ and $w$ decreases as the scalar rolls down the
potential (”cooling” models), which occupy most of
the region $w_0 > - 1, w_a > 0$ (see Barger, Guarnaccia
, Marfatia 2006). These typically arise in models
of dynamical supersymmtery breaking (Binetruy
1999; Masiero, Pietroni, Rosati 2000) and supergravity
(Brax and Martin 1999; Copeland, Nunes, Rosati
2000) including the ”freezing” models in Caldwell \&
Linder (2005) in which the potential has a minimum at
$\phi = \infty$.}
\item{ For ”phantom” models with $w_0 < - 1$ (see
Caldwell 2002), our constraint $w_a \geq -3/4 -(w_0 + 3/2)$
{\it excludes a major portion} of the parameter space corresponding
to models for which the equation of state
crossed the phantom divide line $w = -1$ from a higher
value.}
\end{itemize}

\subsection{Improving constraints with improved measurements.}

For each of the observables we considered, our constraints
can be greatly tightened when improved, reliable
measurements of the actual baryon fraction in
galaxies, and of the relative weight of each baryonic
component, will be available. E.g., a stellar to halo
mass ratio $M_{star}/M = 0.25 f_b$ (a value favored by present
hydrodynamical N-body simulations) would greatly
tighten the constraints from the stellar mass function,
allowing us to rule out all models with $w_a \geq 1$
presently allowed by the distant quasar method. \\

Also, the constraints from the abundance of submm galaxies
at high redshifts could be greatly tightened when
the gas mass of H-dropouts will be reliably measured.
Spectroscopic follow-up of H-dropout galaxies with
future facilities (e.g. the James Webb Space Telescope,
JWST) will add a valuable improvement to the
present analysis. Future measurements on the $H_2$ gas
fraction at high redshift will also allow to reduce the
present gap with respect to the theoretical expectations
$f_{H_2} \approx 0.1 - 0.4$. 

\medskip

Such improved measurements will probably need
the advent of future facilities. E.g., while a more accurate
estimate of the gas-to-stellar mass fraction for the
SPT031158 pair could in principle be inferred from
their stellar mass, the latter is currently poorly constrained:
their rest-frame optical SED is only sampled
by two (IRAC CH1 and CH2 for the Eastern source)
and four (F125W, F160W, IRAC CH1 and CH2 for
the Western source) photometric points, resulting into
a $1-\sigma$ uncertainty on the inferred stellar mass spanning
a factor of $15-20$. In the next future, JWST will
easily improve the accuracy in the stellar mass of the
SPT031158 pair by providing a much more detailed
characterization of the rest-frame optical and near-IR
SED of these galaxies.

\subsection{Statistics}
Increasing the statistics of high-redshift massive
objects will also greatly tighten present constraints (as
shown by the comparison between figs. 2 and 3). Large
surveys from space with the Euclid (Laureijs et al.
2011) and the Wide Field Infrared Survey Telescope
(WFIRST, Spergel et al. 2015) satellites will increase
the number of massive, high-z galaxies by orders of
magnitude with respect to current HST samples. The
Euclid surveys will cover 15000 $deg^2$ at $H \leq 24$ mag
depth, and 40 $deg^2$ at $H \leq 26$, while the WFIRST High
Latitude Survey will observe 2200 $deg^2$ at $H \leq 26.7$.\\

As a reference, the CANDELS GOODS-South sample
comprises only one source with $M_{star} \approx 10^{11} M_{\odot}$
at $z \geq 6$ for $H \leq 24$, and $7$ such sources at $H \leq 26.7$
on an area $\approx 0.05 \,deg^2$. The statistical uncertainty on
the stellar mass function will thus be reduced by a factor
$30-300$ by the aforementioned surveys, extending
also to higher masses than those probed today. Unfortunately,
systematic uncertainties will then dominate
the error budget, mostly because the observed H band
samples the rest-frame UV at $z \geq 6$ resulting in a potentially
biased and incomplete selection of massive
sources. In addition, the lack of information in the
optical rest-frame adds significant uncertainties in the
physical parameters estimated from SED-fitting. This
problem will be overcome by JWST observations with
the Mid-Infrared Instrument at $5.6 -25 \mu m$, albeit on a
much smaller area than Euclid and WFIRST. \\

Despite the lack of any plan for mid-IR large surveys from
space, the combination of H-selected samples from future
cosmological surveys, and improved characterization
of high-z objects on smaller areas thanks to JWST,
will lead to tighter constraints on the high-mass end of
the stellar mass function at $z \geq 6$, and thus on the 
parameter space $(w_0, w_a)$ of DDE models.

\begin{acknowledgements}
We acknowledge support from INAF under PRIN SKA/CTA FORECaST and PRIN SKA-CTA-INAF ASTRI/CTA Data Challenge. 
N.G.S acknowledges CNRS for Emeritus Director of Research contract in LERMA-Observatoire de Paris-PSL-Sorbonne U.
 We thank the referee for helpful and constructive comments that helped to improve the paper.
\end{acknowledgements}

\end{document}